\documentclass[aps, prd, showkeys, showpacs]{revtex4}

\usepackage[small]{caption}
\usepackage{amsmath,amsfonts,amscd,amssymb,epsf, epsfig}\usepackage{graphicx}

\newcommand{\Z}{{\mathbb{Z}}}

\newcommand{\R}{\mathbb{R}}

\newcommand{\pa}{\partial}

\begin{document}

 \title{Finite temperature Casimir effect for  scalar field with Robin boundary conditions in spacetime with extra dimensions }
 \author{L.P. Teo}\email{ LeePeng.Teo@nottingham.edu.my}\address{Department of Applied Mathematics, Faculty of Engineering, University of Nottingham Malaysia Campus, Jalan Broga, 43500, Semenyih, Selangor Darul Ehsan, Malysia. }

\begin{abstract}
In this article, we study the finite temperature Casimir effect for   scalar field with Robin boundary conditions on two parallel plates in a background spacetime that has a compact internal manifold with arbitrary geometry. The finite temperature  Casimir force acting on the plates is derived using the piston approach, with which we show that the  divergences of the Casimir forces from the region between the plates and the region outside the plates cancel each other. The sign and asymptotic behaviors of the Casimir force at different limits such as small and large plate separation, low and high temperature, are studied in detail. The range of Robin coefficients where the Casimir force are always attractive or always repulsive are determined. There are also some range of the Robin coefficients where the Casimir force is shown to be attractive at small plate separation and repulsive at large plate separation, or vise versa.

\end{abstract}

 \maketitle
\section{Introduction}

Although Casimir effect has attracted a lots of attention   due to its important roles in various fields of physics \cite{1}, there have been relatively few works on Casimir effect for quantum fields subject to Robin boundary conditions. Robin boundary conditions arise naturally in the study of Casimir effect for electromagnetic fields in perfectly conducting spheres \cite{2}. However, in parallel plate configuration, perfectly conducting boundary conditions only lead to Dirichlet boundary conditions and Neumann boundary conditions. Nevertheless, it was shown that in some geometries, Robin boundary conditions can be considered as extensions of perfectly conducting boundary conditions, where the Robin coefficients are related to the skin-depth parameter describing the finite penetration of the field into the boundary \cite{3, 4, 5}.  Robin type conditions are needed for conformally invariant field theories in the presence of boundaries, since Robin boundary conditions can be made conformally invariant, but ordinary Neumann boundary conditions cannot. The Casimir effect for fields  with Robin boundary conditions was studied from different perspectives and under different context in \cite{5, 20, 6, 7, 8, 9, 36,19, 18} for parallel plate geometry,  in \cite{26} for a $D$-dimensional ball, in \cite{27} for spherical shell geometry,  and in \cite{25} for the configuration of two spheres. On the other hand, quantum fields with nonlocal boundary conditions can also be reduced to one with Robin boundary conditions \cite{17}.
The importance of the Robin boundary conditions to spacetime models and quantum gravity has been emphasized in \cite{10,11}.  For Randall-Sundrum  spacetime model \cite{12, 13}, Robin boundary conditions are natural for scalar fields and fermion fields, where the Robin coefficients are related to the curvature scale and the boundary mass terms of the fields  \cite{14,15,10_19_1,16}. Casimir effect in braneworld models with Robin boundary conditions on the branes was discussed extensively in \cite{16, 31, 22,23,28,29,30,33,34,35,37,24,56}.

Despite the importance of the Robin boundary conditions in field theory,  the research on Casimir effect with Robin boundary conditions has been limited to a small community, and very few of the works has taken into account the influence of finite temperature.  The goal of this paper is to consider the thermal Casimir effect for the  parallel plate configuration. Since spacetimes with extra dimensions have become prevalent in physics, we are going to discuss along this line. Our setup is the same as in \cite{18}, i.e., we consider a pair of codimension one parallel plates in the background spacetime $M^{d_1+1}\times\mathcal{N}^n$, where $M^{d_1+1}$ is the $(d_1+1)$-dimensional Minskowski spacetime, and $\mathcal{N}^n$ is a compact internal space. In fact, Casimir effect on parallel plates in a background spacetime of this form has been investigated in \cite{38,39,40,41,42,43,44}, when the boundary conditions are either Dirichlet boundary conditions or Neumann boundary conditions in the case of scalar fields, and perfect conductor boundary conditions in the case of electromagnetic fields. In the cases considered in \cite{38,39,40,41,42}, the dimension of the macroscopic space is  $d_1=3$ and the internal space is assumed to be the simplest one-dimensional compact space $S^1$ or the more general $n$-dimensional toroidal manifold $T^n$. In \cite{43,44}, the internal space is generalized to arbitrary compact manifold and the parallel plates are allowed to be finite with arbitrary cross section. The temperature corrections to the Casimir effect in the scenario of \cite{43,44} were considered in \cite{45,46,47}. It has been shown that for massless scalar field with Dirichlet boundary conditions on both plates or Neumann boundary conditions on both plates, the Casimir force is always attractive, at any temperature and for any spacetime geometry. For massless scalar field with Dirichlet boundary condition on one plate and Neumann boundary condition on another plate, the Casimir force is always repulsive. The present work can be considered as a generalization of \cite{18} to take into account the finite temperature correction, or a generalization of \cite{47} where the boundary conditions are the more general Robin boundary conditions instead of the Dirichlet boundary conditions or Neumann boundary conditions. We are mainly interested in studying the dependence of the Casimir force on the interplay between the spacetime geometry, temperature and boundary conditions.

\begin{figure}
\epsfxsize=0.5\linewidth \epsffile{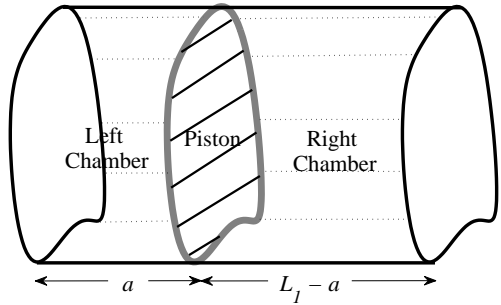} \caption{\label{f1} A movable piston inside a closed cylinder divides the cylinder into two chambers.}\end{figure}

As in \cite{44,46,47}, we first consider the Casimir force acting on a piston which moves freely inside a closed cylinder with arbitrary cross section (see FIG. \ref{f1}). In Section \ref{ce}, the Casimir energy in the two regions   divided by piston is computed using cut-off method. In Section \ref{cf}, the Casimir force acting on the piston is derived. The limit where one end of the cylinder is moved to infinity is equivalent to two parallel plates embedded orthogonally inside an infinitely long cylinder. Letting the cross section of the plates become infinitely large give the usual infinite parallel plate configuration. This piston scenario introduced by Cavalcanti \cite{48} has the advantage that the contribution to the Casimir force due to the vacuum fluctuations  of the quantum fields in the region between the plates and the region outside the plates are both taken into account, and the divergences of the Casimir forces from the two regions will   cancel. In Section \ref{acf}, we study in detail the sign of the Casimir force at small and large plate separations.

In this article, we use the units where $\hbar=c=k_B=1$.

\section{The Casimir energy}\label{ce}
As in \cite{18, 47}, we consider a massive scalar field $\varphi(x)$  in a background spacetime of the form $M^{d_1+1}\times \mathcal{N}^n$, where $M^{d_1+1}$ is the $(d_1+1)$-dimensional Minkowski spacetime and $\mathcal{N}^n$ is an $n$-dimensional compact internal manifold. Let \begin{equation*}\begin{split}
ds^2 =& g_{\mu\nu} dx^{\mu}dx^{\nu} =\eta_{\alpha\beta}dx^{\alpha}dx^{\beta}-G_{ab}dy^a dy^b,\\
\mu,\nu=&0,1,\ldots,d;\;\alpha,\beta=0,1,\ldots, d_1; \;a,b=1,\ldots, n,\end{split}
\end{equation*} be the spacetime metric,   where  $\eta_{\alpha\beta}=\text{diag}\,(1, -1,\ldots, -1)$,  $y^{a} = x^{d_1+a}$ for $a=1,\ldots,n $ and $G_{ab} dy^ady^b$ is a Riemannian metric on $\mathcal{N}^n$, and $d=d_1+n$. The equation of motion for a massive scalar  field $\varphi(x)$ is
\begin{equation*}
\left(\frac{1}{\sqrt{|g|}}\pa_{\mu}\sqrt{|g|}g^{\mu\nu}\pa_{\nu} +m^2  \right)\varphi(x)=0.
\end{equation*} We will first consider the  Casimir force acting on a movable piston inside a closed cylinder of arbitrary cross section as shown in FIG. \ref{f1}. Let the cylinder be the region $[0,L_1]\times \Omega\times \mathcal{N}^n$ in $\R^{d_1+n}$, where $\Omega$ is the cross section of the piston, which we assume to be a simply connected region in $\R^{d_1-1}$.  The position of the piston is denoted by $x^1=a$. If the right chamber of the cylinder is  infinitely long, i.e., $L_1\rightarrow \infty$, we  obtain  the configuration equivalent to two codimension one parallel plates located at $x^1=0$ and $x^1=a$ embedded in an infinitely long cylinder.  For the boundary conditions, we assume that the field $\varphi(x)$ satisfies Dirichlet boundary conditions on the curved surface $[0,L_1]\times\pa\Omega\times\mathcal{N}^n$ of the cylinder.   The interesting  part is the boundary conditions on the three plates perpendicular to the $x^1$-direction, which we take to be Robin boundary conditions:
\begin{equation}\label{eq7_22_1}
\left.\left(\alpha_1-\beta_1\frac{\pa}{\pa x^1}\right)\varphi(x^1) \right|_{x^1=0}=0, \hspace{0.5cm}\left.\left(\alpha_2+\beta_2\frac{\pa}{\pa x^1}\right)\varphi(x^1) \right|_{x^1=a}=0, \hspace{0.5cm}\left.\left(\alpha_3-\beta_3\frac{\pa}{\pa x^1}\right)\varphi(x^1) \right|_{x^1=L_1}=0,
\end{equation}   where $\alpha_1,\alpha_2,\alpha_3$ are nonnegative dimensionless constants, and $\beta_1,\beta_2,\beta_3$ are constants with dimension of length.  For $i=1,2,3$, $\alpha_i$ and $\beta_i$ cannot be zero simultaneously.  $\beta_i=0$ corresponds to the Dirichlet boundary conditions,  $\alpha_i=0$ corresponds to the Neumann boundary conditions, and the general Robin boundary conditions only depend on the ratio $\beta_i/\alpha_i$. We would like to remark that our $\beta_i/\alpha_i$ is equal to the $-\beta_i$ in \cite{18}.

To find the Casimir force acting on the piston, one does not have to calculate the Casimir energy of the region outside  the cylinder, since it is independent of the position of the piston \cite{48}. Therefore, we only have to compute the Casimir energies inside the left chamber and the right chamber.
It is easy to see that the Casimir energy inside the right chamber can be obtained from the Casimir energy inside the left chamber by replacing $a$ with $L_1-a$, $\alpha_1$ with $\alpha_3$ and $\beta_1$ with $\beta_3$.

Using separation of variables, one can show that the eigenmodes of the field $\varphi(x)$ confined in the left chamber satisfying the boundary conditions specified above are given by
\begin{equation}\label{eq3_27_2}
\varphi_{k,j,l}(x) = e^{-i\omega t}\left(A\sin z_kx^1+B\cos z_kx^1\right) \phi_{\Omega,j}\left(x^2,\ldots, x^{d_1}\right) \phi_{\mathcal{N},l}(y). \hspace{1cm} j\in\mathbb{N}, l\in\mathbb{N}_0=\mathbb{N}\cup\{0\}.
\end{equation}For $j=1,2,\ldots, $, $\phi_{\Omega,j}(x^2, \ldots, x^{d_1})$ is an eigenfunction with eigenvalue $\omega_{\Omega,j}^2>0$ for the Laplace operator with Dirichlet boundary conditions on $\pa\Omega$. For $l=0,1,2,\ldots$, $\phi_{\mathcal{N},l}(y)$ is an eigenfunction  with eigenvalue $\omega_{\mathcal{N},l}^2$ for the Laplace operator   on $\mathcal{N}^n$. By convention, when $l=0$, $\phi_{\mathcal{N},0}(y)$ is a constant function with eigenvalue $\omega_{\mathcal{N},0}^2=0$.  The boundary conditions \eqref{eq7_22_1} imposed on $x^1=0$ and $x^1=a$ imply that
\begin{equation}\label{eq7_22_2}\begin{split}
&\hspace{1cm}- \beta_1 z_kA +\alpha_1B =0,\\
&\left(\alpha_2\sin (az_k)+ \beta_2z_k \cos (az_k) \right) A+\left(\alpha_2 \cos (az_k)- \beta_2z_k \sin (az_k)\right)B =0.
\end{split}
\end{equation}This system gives a nontrivial solution for $(A,B)$ if and only if  $z_k$  satisfies the following condition:
\begin{equation}\label{eq3_27_3}
F(z) = (\alpha_1\alpha_2-\beta_1\beta_2z^2)\sin  az  +(\alpha_2\beta_1+\alpha_1\beta_2) z\cos az =0.
\end{equation}The zeros of $F(z)$ are either real or purely imaginary. As is shown in  \cite{6}, these zeros are all simple. In the following, we assume that $F(z)$ does not have imaginary zeros. This is the case if \cite{6}:
\begin{equation}\label{eq7_23_10}\begin{split}
\left\{\alpha_1=0,   \beta_2\geq 0\right\} \hspace{0.2cm}\text{or}\hspace{0.2cm}\left\{\alpha_2=0,\beta_1\geq 0\right\}\hspace{0.2cm}\text{or}\hspace{0.2cm}\left\{ \alpha_1\alpha_2\neq 0, \beta_1\geq 0, \beta_2\geq 0\right\} \hspace{0.2cm}\text{or}\hspace{0.2cm}\left\{\alpha_1\alpha_2\neq 0, \frac{\beta_1}{\alpha_1}+\frac{\beta_2}{\alpha_2}+a\leq 0, \;\; \beta_1\beta_2\leq 0\right\}.
\end{split}\end{equation}
 Notice that $z=0$ is always a solution of \eqref{eq3_27_3}. However, it is easy to check that $z_k=0$ gives to a nontrivial function $A\sin z_k x^1+B\cos z_kx^1$ satisfying \eqref{eq7_22_2} if and only if $\alpha_1=\alpha_2=0$, i.e., in the case of Neumann boundary conditions on both plates. We let $0< z_1<z_2< z_3< \ldots$ be all the positive solutions of \eqref{eq3_27_3}. Then the set of eigenfrequencies $\{\omega_{k,j,l}\}$ for the field $\varphi(x)$ is given by
\begin{equation}\label{eq9_17_1}
\omega_{k,j,l}=\sqrt{ z_k^2 + \omega_{\Omega,j}^2+\omega_{\mathcal{N},l}^2+m^2}=\sqrt{ z_k^2 + m_{j,l}^2}, \hspace{0.5cm}k,j\in\mathbb{N}, \;\;l\in\mathbb{N}_0,
\end{equation}where $$m_{j,l}^2=\omega_{\Omega,j}^2+\omega_{\mathcal{N},l}^2+m^2.$$In the case $\alpha_1=\alpha_2=0$, we have to let $k$ starts from $0$ instead of $1$, and $z_0=0$.

The finite temperature Casimir energy inside the left chamber is given by
\begin{equation}\label{eq3_27_5}
\begin{split}
E_{\text{Cas}}^{L}(\lambda)=\sum_{k=1}^{\infty} \sum_{j=1}^{\infty}\sum_{l=0}^{\infty}\left\{\frac{1}{2}\omega_{k,j,l}e^{-\lambda\omega_{k,j,l}} +T\log\left( 1-e^{-\omega_{k,j,l}/T}\right)\right\},
\end{split}
\end{equation}where $\lambda$ is a cut-off parameter.
As was shown in \cite{46,47}, up to the term of order $\lambda^0$,
\begin{equation}\label{eq4_1_3}\begin{split}
E_{\text{Cas}}^{L}(\lambda)=&\sum_{i=0}^{d-1} \frac{\Gamma\left(d+1-i\right)}{\Gamma\left(\frac{d-i}{2}\right)} c_{\text{cyl},i}\lambda^{i-d-1}+\frac{ \log[\lambda\mu]-\psi(1)-\log 2+1}{2\sqrt{\pi}}c_{\text{cyl},d+1}-\frac{T}{2}\left(\zeta_{\text{cyl}, T}'(0)+\log[\mu^2] \zeta_{\text{cyl}, T}(0)\right),\end{split}
\end{equation}where $\mu$ is a normalization constant with dimension length$^{-1}$, $\zeta_{\text{cyl}, T}(s)$ is the
  finite temperature zeta function defined by
\begin{equation*}
\begin{split}
\zeta_{\text{cyl},T}(s)=\sum_{k=1}^{\infty}\sum_{j=1}^{\infty}\sum_{l=0}^{\infty}\sum_{p=-\infty}^{\infty}\left(z_k^2+m_{j,l}^2+[2\pi pT]^2\right)^{-s},
\end{split}
\end{equation*}
and $c_{\text{cyl},i}$, $0\leq i\leq d+1$, are heat kernel coefficients defined by
\begin{equation}\label{eq4_1_4}
\sum_{k=1}^{\infty}\sum_{j=1}^{\infty}\sum_{l=0}^{\infty} e^{-t\left(z_k^2+m_{j,l}^2 \right)}\sim \sum_{i=0}^{d+1}c_{\text{cyl},i}t^{\frac{i-d}{2}}+O\left(t \right)\hspace{1cm}
\text{as}\;\;t\rightarrow 0^+.
\end{equation}It can be shown that (see Appendix \ref{a3}) these coefficients are linear functions of $a$. Moreover, the coefficient of $a$ in $c_{\text{cyl},i}$  is independent of the Robin coefficients $\alpha_i,\beta_i$, $i=1,2$.

Using a more general form of the generalized Abel-Plana formula \cite{49,50} (see Appendix \ref{a1}), we compute $\zeta_{\text{cyl},T}(0)$ and $\zeta_{\text{cyl},T}'(0)$ in Appendix \ref{a2}. The results substituted into \eqref{eq4_1_3} give the Casimir energy in the left chamber:
\begin{equation}\label{eq7_23_8_1}
\begin{split}
E_{\text{Cas}}^{L}(\lambda)=&\Xi_0^L(\lambda)+a\Xi_1(\lambda)\\& +\frac{T}{2}\sum_{j=1}^{\infty}\sum_{l=0}^{\infty}\sum_{p=-\infty}^{\infty}\log\left(
1-\frac{\left( \beta_1 \sqrt{m_{j,l}^2+[2\pi p T]^2}-\alpha_1\right)\left( \beta_2\sqrt{m_{j,l}^2+[2\pi p T]^2}-\alpha_2\right)}{\left( \beta_1 \sqrt{m_{j,l}^2+[2\pi p T]^2}+\alpha_1\right)\left( \beta_2 \sqrt{m_{j,l}^2+[2\pi p T]^2}+\alpha_2\right)}e^{-2a\sqrt{m_{j,l}^2+[2\pi p T]^2}}\right),
\end{split}
\end{equation}where $\Xi_0^L(\lambda)$ and $\Xi_1(\lambda)$ are terms that are independent of $a$, and $\Xi_1(\lambda)$ is also independent of the Robin coefficients. In fact, $\Xi_1(\lambda)$ can be interpreted as the vacuum energy per unit length in the $x^1$ direction that would present in the region between the plates if the plates are absent.

 Our assumption that $F(z)$ does not have imaginary zeros will make each term under the logarithm in \eqref{eq7_23_8_1} nonzero. However, if $\beta_1\alpha_1<0$ or $\beta_2\alpha_2<0$, we may get terms that are infinite. To avoid this kind of complications, from now on we only consider the case where $\beta_1\geq 0$ and $\beta_2\geq 0$.

\section{The Casimir force}\label{cf} For the piston system shown in FIG. \ref{f1}, the Casimir force acting on the piston is given by
\begin{equation*}
\begin{split}
F^{\text{piston}}_{\text{Cas}}(a,L_1; \boldsymbol{\alpha}, \boldsymbol{\beta})=-\frac{\pa}{\pa a}\left( E_{\text{Cas}}^{L}(\lambda)+E_{\text{Cas}}^R(\lambda)\right).
\end{split}
\end{equation*}Upon differentiation with respect to $a$, the terms $\Xi_0^L(\lambda)$ and $\Xi_0^R(\lambda)$ that are independent of $a$ will be killed. On the other hand, for the term proportional to $a$ in $E_{\text{Cas}}^{L}(\lambda)+E_{\text{Cas}}^R(\lambda)$, we find that it is equal to $a\Xi_1+(L_1-a)\Xi_1=L_1\Xi_1$, which is independent of $a$. Therefore, this term will also be killed after differentiation with respect to $a$. Consequently, all the terms that would diverge when $\lambda\rightarrow 0^+$ will not contribute to the Casimir force. Therefore, we can set $\lambda=0$ after taking derivative with respect to $a$ to obtain
\begin{equation}\label{eq7_29_4}
F^{\text{piston}}_{\text{Cas}}(a,L_1; \boldsymbol{\alpha}, \boldsymbol{\beta})=-\lim_{\lambda\rightarrow 0^+}\frac{\pa}{\pa a}\left( E_{\text{Cas}}^{L}(\lambda)+E_{\text{Cas}}^R(\lambda)\right)=F^{L}_{\text{Cas}}(a;\alpha_1,\beta_1,\alpha_2,\beta_2)-F_{\text{Cas}}^R(a,L_1;\alpha_3,\beta_3,\alpha_2,\beta_2),
\end{equation}
where
\begin{equation}\label{eq7_23_11}
F^{L}_{\text{Cas}}(a;\alpha_1,\beta_1,\alpha_2,\beta_2)=-T\sum_{j=1}^{\infty}\sum_{l=0}^{\infty}\sum_{p=-\infty}^{\infty}\frac{\sqrt{m_{j,l}^2+[2\pi p T]^2}}{\frac{\left( \beta_1 \sqrt{m_{j,l}^2+[2\pi p T]^2}+\alpha_1\right)\left( \beta_2 \sqrt{m_{j,l}^2+[2\pi p T]^2}+\alpha_2\right)}{\left( \beta_1 \sqrt{m_{j,l}^2+[2\pi p T]^2}-\alpha_1\right)\left( \beta_2\sqrt{m_{j,l}^2+[2\pi p T]^2}-\alpha_2\right)}e^{2a\sqrt{m_{j,l}^2+[2\pi p T]^2}}-1},
\end{equation}and $$F_{\text{Cas}}^R(a,L_1;\alpha_3,\beta_3,\alpha_2,\beta_2)=F_{\text{Cas}}^L(L_1-a;\alpha_3,\beta_3,\alpha_2,\beta_2).$$
Notice that   even though different Robin conditions are imposed on the walls $x^1=0$ and $x^1=L_1$, the divergent terms of the Casimir energy will not contribute to the Casimir force acting on the piston. Therefore the Casimir force acting on the piston is independent of the regularization procedure, as in the piston system for scalar fields with Dirichlet or Neumann boundary conditions. A  crucial reason why this is so is that the term $\Xi_1(\lambda)$  is independent of the Robin coefficients.

It is easy to see that $F^R_{\text{Cas}}(a,L_1;\alpha_3,\beta_3,\alpha_2,\beta_2)$ vanishes as $L_1\rightarrow \infty$. In other words, $F^{L}_{\text{Cas}}(a;\alpha_1,\beta_1,\alpha_2,\beta_2)$ can be interpreted as the Casimir force acting between two parallel plates embedded orthogonally inside an infinitely long cylinder. In the following, we will drop the superscript $L$ when we discuss the Casimir force between parallel plates. Setting $\beta_1=\beta_2=0$ or $\alpha_1=\alpha_2=0$ or $\beta_1=\alpha_2=0$ in \eqref{eq7_23_11}, one obtains respectively the Casimir force acting on a pair of parallel plates with Dirichlet-Dirichlet, Neumann-Neumann or Dirichlet-Neumann boundary conditions. The results coincide with the results derived in \cite{47}.

Notice that the Casimir force between   parallel plates $F_{\text{Cas}}(a;\alpha_1,\beta_1,\alpha_2,\beta_2)$ \eqref{eq7_23_11} comes from the derivative with respect to the third term
$$\Delta E_{\text{Cas}}=\frac{T}{2}\sum_{j=1}^{\infty}\sum_{l=0}^{\infty}\sum_{p=-\infty}^{\infty}\log\left(
1-\frac{\left( \beta_1 \sqrt{m_{j,l}^2+[2\pi p T]^2}-\alpha_1\right)\left( \beta_2\sqrt{m_{j,l}^2+[2\pi p T]^2}-\alpha_2\right)}{\left( \beta_1 \sqrt{m_{j,l}^2+[2\pi p T]^2}+\alpha_1\right)\left( \beta_2 \sqrt{m_{j,l}^2+[2\pi p T]^2}+\alpha_2\right)}e^{-2a\sqrt{m_{j,l}^2+[2\pi p T]^2}}\right)
$$in \eqref{eq7_23_8_1}, which was called the interaction term \cite{18}. Usually this is also regarded as the renormalized Casimir energy between the plates. It has the property that it vanishes when the plate separation $a$ goes to infinity.

 The expression for the Casimir force between parallel plates \eqref{eq7_23_11} shows that the Casimir force is always attractive (negative) if the same boundary conditions (i.e., $\beta_1/\alpha_1=\beta_2/\alpha_2$) are imposed on both the plates. This is a special case of the theorem \cite{51,52} which states that the Casimir force between two bodies with the same property is attractive. From \eqref{eq7_23_11}, we also find that the Casimir force decays exponentially as the   separation of the plates $a$ becomes large.  As a function of the temperature $T$,  \eqref{eq7_23_11}   shows that the high temperature leading term of the Casimir force is linear in $T$, given by the sum of the terms with $p=0$.
 On the other hand, let $r=(\text{Vol}(\Omega))^{1/(d_1-1)}$ and $R=(\text{Vol}(\mathcal{N}^n))^{1/n}$ be the size of the cross section $\Omega$ and the size of the internal space $\mathcal{N}^n$ respectively.
  Recall that $m_{j,l}^2=\omega_{\Omega,j}^2+\omega_{\mathcal{N},l}^2+m^2$. Since $\omega_{\Omega,j} \propto 1/r$ and $\omega_{\mathcal{N},l}\propto 1/R$,  \eqref{eq7_23_11} shows that the Casimir force $F_{\text{Cas}}(a;\alpha_1,\beta_1,\alpha_2,\beta_2)$ between a pair of parallel plates goes to zero when the size $r$ of the cross section goes to zero or the mass $m$ goes to infinity. When the size $R$ of the internal space goes to zero, all the terms with $l\neq 0$ go to zero, and the limit of the Casimir force is the Casimir force in the $(d_1+1)$-dimensional Minkowski spacetime given by the sum of the terms with $l=0$.

  For the low temperature behavior, we use the   Abel-Plana summation formula \eqref{eq7_27_1} with
\begin{equation*}
f(z)=-2T\sum_{j=1}^{\infty}\sum_{l=0}^{\infty} \frac{\sqrt{m_{j,l}^2+[2\pi   Tz]^2}}{\frac{\left( \beta_1 \sqrt{m_{j,l}^2+[2\pi  Tz]^2}+\alpha_1\right)\left( \beta_2 \sqrt{m_{j,l}^2+[2\pi   Tz]^2}+\alpha_2\right)}{\left( \beta_1 \sqrt{m_{j,l}^2+[2\pi   Tz]^2}-\alpha_1\right)\left( \beta_2\sqrt{m_{j,l}^2+[2\pi   Tz]^2}-\alpha_2\right)}e^{2a\sqrt{m_{j,l}^2+[2\pi  T z]^2}}-1}.
\end{equation*}
The term
$$\int_0^{\infty}f(x)dx$$ on the right-hand side of \eqref{eq7_27_1} gives the zero temperature Casimir force acting on the parallel plates:
\begin{equation}\label{eq7_27_6}
\begin{split}
F^{T=0}_{\text{Cas}}(a;\alpha_1,\beta_1,\alpha_2,\beta_2)=&-\frac{1}{\pi}\sum_{j=1}^{\infty}\sum_{l=0}^{\infty}\int_0^{\infty}\frac{\sqrt{x^2+m_{j,l}^2}}{\frac{\left( \beta_1 \sqrt{x^2+m_{j,l}^2 }+\alpha_1\right)\left( \beta_2 \sqrt{x^2+m_{j,l}^2 }+\alpha_2\right)}{\left( \beta_1 \sqrt{x^2+m_{j,l}^2 }-\alpha_1\right)\left( \beta_2\sqrt{x^2+m_{j,l}^2 }-\alpha_2\right)}e^{2a\sqrt{x^2+m_{j,l}^2}}-1}dx\\
=&-\frac{1}{\pi}\sum_{j=1}^{\infty}\sum_{l=0}^{\infty}\int_{m_{j,l}}^{\infty}\frac{x^2}{\frac{\left( \beta_1 x+\alpha_1\right)\left( \beta_2 x+\alpha_2\right)}{\left( \beta_1x-\alpha_1\right)\left( \beta_2x-\alpha_2\right)}e^{2ax}-1}\frac{dx}{\sqrt{x^2-m_{j,l}^2}}.
\end{split}
\end{equation}The term
\begin{equation*}
i\int_0^{\infty}\frac{f(iy)-f(-iy)}{e^{2\pi y}-1}dy
\end{equation*}gives
\begin{equation*}
\begin{split}
-\frac{1}{\pi}\sum_{j=1}^{\infty}\sum_{l=0}^{\infty}\int_{m_{j,l}}^{\infty}\frac{\sqrt{u^2-m_{j,l}^2}}{e^{u/T}-1}du=-\frac{T}{\pi}\sum_{j=1}^{\infty}\sum_{l=0}^{\infty}\sum_{p=1}^{\infty}
\frac{m_{j,l}}{p}K_1\left(\frac{p m_{j,l}}{T}\right).
\end{split}
\end{equation*}This term is independent of $a$ and  the Robin coefficients $\alpha_i, \beta_i$, $i=1,2$. Finally, it is easy to check that the poles of $f(z)$ are exactly at \begin{equation*}
\begin{split}
z=\pm \frac{i}{2\pi T}\sqrt{ z_k^2+m_{j,l}^2},\hspace{1cm} k=1,2,\ldots,j=1,2,\ldots, l=0,1,2,\ldots,
\end{split}
\end{equation*}with residues
\begin{equation*}
\text{Res}_{z=\pm \frac{i}{2\pi T}\sqrt{ z_k^2+m_{j,l}^2}}f(z) =\mp\frac{i}{2\pi}\frac{z_k^2}{\sqrt{z_k^2+m_{j,l}^2}\left(a+\frac{\alpha_1\beta_1}{\beta_1^2z_k^2+\alpha_1^2}+\frac{\alpha_2\beta_2}{\beta_2^2z_k^2+\alpha_2^2}\right)}.
\end{equation*}Therefore,
\begin{equation*}
\begin{split}
\pi i \sum_{ y>0} \frac{\text{Res}_{z=iy} f(z)-\text{Res}_{z=-iy}f(z)}{ e^{2\pi y}-1}=\sum_{j=1}^{\infty}\sum_{l=0}^{\infty}\sum_{k=1}^{\infty}
\frac{z_k^2}{\sqrt{z_k^2+m_{j,l}^2}\left(a+\frac{\alpha_1\beta_1}{\beta_1^2z_k^2+\alpha_1^2}+\frac{\alpha_2\beta_2}{\beta_2^2z_k^2+\alpha_2^2}\right)}
\frac{1}{\exp\left(\frac{\sqrt{z_k^2+m_{j,l}^2}}{T}\right)-1}.
\end{split}
\end{equation*}From these, we find that the temperature correction to the Casimir force is
\begin{equation}\label{eq7_27_2}
\begin{split}
\Delta_TF_{\text{Cas}}(a;\alpha_1,\beta_1,\alpha_2,\beta_2)=&-\frac{T}{\pi}\sum_{j=1}^{\infty}\sum_{l=0}^{\infty}\sum_{p=1}^{\infty}
\frac{m_{j,l}}{p}K_1\left(\frac{p m_{j,l}}{T}\right)\\&+\sum_{j=1}^{\infty}\sum_{l=0}^{\infty}\sum_{k=1}^{\infty}
\frac{z_k^2}{\sqrt{z_k^2+m_{j,l}^2}\left(a+\frac{\alpha_1\beta_1}{\beta_1^2z_k^2+\alpha_1^2}+\frac{\alpha_2\beta_2}{\beta_2^2z_k^2+\alpha_2^2}\right)}
\frac{1}{\exp\left(\frac{\sqrt{z_k^2+m_{j,l}^2}}{T}\right)-1}.
\end{split}
\end{equation}In the case of $d_1=3$ with Dirichlet-Dirichlet boundary conditions, $\beta_1=\beta_2=0$ and $z_k=\pi k/a$, we find that this formula agrees with the formula (3.7) we derived in \cite{46}. From \eqref{eq7_27_2}, it is obvious that the temperature correction  goes to zero exponentially fast when the temperature $T$ approaches zero. When $a\ll 1/T$, since $z_k\propto 1/a$, the leading term of the temperature correction is the first term on the right hand side of  \eqref{eq7_27_2} that is independent of $a$.

The next thing we would like to investigate is the limit of the Casimir force when the cross section of the plates is infinitely large, i.e. $\text{Vol}(\Omega)\rightarrow \infty$. In this case, we have to consider the Casimir force density acting on the plates, which is defined as
\begin{equation*}\begin{split}
\mathcal{F}_{\text{Cas}}(a;\alpha_1,\beta_1,\alpha_2,\beta_2)=&\lim_{r\rightarrow \infty} \frac{F_{\text{Cas}}(a;\alpha_1,\beta_1,\alpha_2,\beta_2)}{\text{Vol}(\Omega)}.
\end{split}\end{equation*}This limit is given by replacing $\omega_{\Omega,j}$ by $\omega$ and turning the summation over $j\in \mathbb{N}$ into an integral:
\begin{equation}\label{eq7_27_5}
\frac{1}{\text{Vol}(\Omega)}\sum_{j=1}^{\infty} g(\omega_{\Omega, j} )\xrightarrow {\text{Vol}(\Omega)\rightarrow\infty}\frac{1}{ 2^{d_1-2}\pi^{\frac{d_1-1}{2}}\Gamma\left(\frac{d_1-1}{2}\right)}\int_0^{\infty}\omega^{d_1-2} g(\omega)d\omega.
\end{equation}After a change of variables, we find that the finite temperature Casimir force density acting on a pair of infinite parallel plates is given by
\begin{equation}\label{eq7_27_4}
\begin{split}
\mathcal{F}_{\text{Cas}}(a;\alpha_1,\beta_1,\alpha_2,\beta_2)=-\frac{T}{2^{d_1-2}\pi^{\frac{d_1-1}{2}}\Gamma\left(\frac{d_1-1}{2}\right)}\sum_{l=0}^{\infty}\sum_{p=-\infty}^{\infty}\int_{\sqrt{m_l^2+[2\pi p T]^2}}^{\infty}\frac{\left(x^2- m_l^2-[2\pi p T]^2\right)^{\frac{d_1-3}{2}}x^2dx}{\frac{(\beta_1 x+\alpha_1)(\beta_2x+\alpha_2)}{(\beta_1 x-\alpha_1)(\beta_2x-\alpha_2)}e^{2ax}-1},
\end{split}
\end{equation}where $$m_l=\sqrt{\omega_{\mathcal{N},l}^2+m^2}.$$The corresponding interaction term of the Casimir energy density is
\begin{equation}\label{eq7_27_4_1}
\begin{split}
\Delta\mathcal{E}_{\text{Cas}}(a;\alpha_1,\beta_1,\alpha_2,\beta_2)=&\frac{T}{2^{d_1-1}\pi^{\frac{d_1-1}{2}}\Gamma\left(\frac{d_1-1}{2}\right)}\sum_{l=0}^{\infty}\sum_{p=-\infty}^{\infty}\int_{\sqrt{m_l^2+[2\pi p T]^2}}^{\infty} \left(x^2- m_l^2-[2\pi p T]^2\right)^{\frac{d_1-3}{2}}x\\&\hspace{3cm}\times\log\left\{1-\frac{(\beta_1 x-\alpha_1)(\beta_2x-\alpha_2)}{(\beta_1 x+\alpha_1)(\beta_2x+\alpha_2)}e^{-2ax}\right\}dx.
\end{split}
\end{equation}
One can check that in the case of Dirichlet-Dirichlet, Neumann-Neumann or Dirichlet-Neumann boundary conditions, \eqref{eq7_27_4} agrees with the results derived in \cite{47}.

Applying the prescription \eqref{eq7_27_5} to the zero temperature Casimir force \eqref{eq7_27_6}, we find that the zero temperature Casimir force density acting on a pair of infinite parallel plates is
\begin{equation}\label{eq7_28_1}
\begin{split}
&\mathcal{F}_{\text{Cas}}^{T=0}(a;\alpha_1,\beta_1,\alpha_2,\beta_2)\\=&-\frac{1}{2^{d_1-2}\pi^{\frac{d_1+1}{2}}\Gamma\left(\frac{d_1-1}{2}\right)}\sum_{l=0}^{\infty}\int_0^{\infty} \omega^{d_1-2}\int_{\sqrt{\omega^2+m_l^2}}^{\infty}\frac{x^2}{\frac{\left( \beta_1 x+\alpha_1\right)\left( \beta_2 x+\alpha_2\right)}{\left( \beta_1x-\alpha_1\right)\left( \beta_2x-\alpha_2\right)}e^{2ax}-1}\frac{dx}{\sqrt{x^2-\omega^2- m_l^2}}d\omega\\
=&-\frac{1}{2^{d_1-1}\pi^{\frac{d_1}{2}}\Gamma\left(\frac{d_1}{2}\right)}\sum_{l=0}^{\infty}\int_{m_l}^{\infty}\frac{x^2(x^2-m_l^2)^{\frac{d_1-2}{2}}}{\frac{\left( \beta_1 x+\alpha_1\right)\left( \beta_2 x+\alpha_2\right)}{\left( \beta_1x-\alpha_1\right)\left( \beta_2x-\alpha_2\right)}e^{2ax}-1}dx.
\end{split}
\end{equation}This agrees with the result of \cite{18}. The finite temperature Casimir force density is the sum of the zero temperature Casimir force density \eqref{eq7_28_1} and the thermal correction $\Delta_T\mathcal{F}_{\text{Cas}}$ given by
\begin{equation}\label{eq7_28_2}
\begin{split}
&\Delta_T\mathcal{F}_{\text{Cas}}(a;\alpha_1,\beta_1,\alpha_2,\beta_2)=-\frac{T}{ 2^{d_1-2}\pi^{\frac{d_1+1}{2}}\Gamma\left(\frac{d_1-1}{2}\right)}\sum_{l=0}^{\infty}\sum_{p=1}^{\infty}\int_0^{\infty}\omega^{d_1-2}
\frac{\sqrt{\omega^2+m_l^2}}{p}K_1\left(\frac{p \sqrt{\omega^2+m_l^2}}{T}\right)d\omega\\
&+\frac{1}{ 2^{d_1-2}\pi^{\frac{d_1-1}{2}}\Gamma\left(\frac{d_1-1}{2}\right)}\sum_{l=0}^{\infty}\sum_{k=1}^{\infty}\int_0^{\infty}\omega^{d_1-2}
\frac{z_k^2}{\sqrt{z_k^2+\omega^2+m_{l}^2}\left(a+\frac{\alpha_1\beta_1}{\beta_1^2z_k^2+\alpha_1^2}+\frac{\alpha_2\beta_2}{\beta_2^2z_k^2+\alpha_2^2}\right)}
\frac{1}{\exp\left(\frac{\sqrt{z_k^2+\omega^2+m_{l}^2}}{T}\right)-1}d\omega\\
=&-\frac{T^{\frac{d_1+1}{2}}}{2^{\frac{d_1-1}{2}}\pi^{\frac{d_1+1}{2}}}\sum_{l=0}^{\infty}\sum_{p=1}^{\infty}\left(\frac{m_l}{p}\right)^{\frac{d_1+1}{2}}K_{\frac{d_1+1}{2}}\left(\frac{p m_l}{T}\right)\\&+\frac{T^{\frac{d_1-2}{2}}}{2^{\frac{d_1-2}{2}}\pi^{\frac{d_1}{2}}}\sum_{l=0}^{\infty}\sum_{k=1}^{\infty}\sum_{p=1}^{\infty}\frac{z_k^2}
{\left(a+\frac{\alpha_1\beta_1}{\beta_1^2z_k^2+\alpha_1^2}+\frac{\alpha_2\beta_2}{\beta_2^2z_k^2+\alpha_2^2}\right)}\left(\frac{\sqrt{z_k^2+m_l^2}}{p}\right)^{\frac{d_1-2}{2}}
K_{\frac{d_1-2}{2}}\left(\frac{p}{T}\sqrt{z_k^2+m_l^2}\right).
\end{split}
\end{equation}In the massless case, the sum of the terms with $l=0$ in the first term on the right hand side of \eqref{eq7_28_2} has to be replaced by
\begin{equation*}
\lim_{m\rightarrow 0^+}\left\{-\frac{T^{\frac{d_1+1}{2}}}{2^{\frac{d_1-1}{2}}\pi^{\frac{d_1+1}{2}}} \sum_{p=1}^{\infty}\left(\frac{m}{p}\right)^{\frac{d_1+1}{2}}K_{\frac{d_1+1}{2}}\left(\frac{p m}{T}\right)\right\}=-\frac{\Gamma\left(\frac{d_1+1}{2}\right)\zeta_R(d_1+1)}{\pi^{\frac{d_1+1}{2}}}T^{d_1+1}.
\end{equation*}We observe that for massive case, the temperature correction term to the Casimir force density acting on a pair of infinite parallel plates is exponentially suppressed. However, for the massless case, the leading term of the thermal correction is of order $T^{d_1+1}$, and this leading term is independent of the boundary conditions imposed on the plates.

The low temperature expansion of the interaction term of the Casimir energy density can be obtained by directly integrating \eqref{eq7_28_1} and \eqref{eq7_28_2}. It is given by
\begin{equation*}
\begin{split}
\Delta\mathcal{E}_{\text{Cas}}(a;\alpha_1,\beta_1,\alpha_2,\beta_2)=&\frac{1}{2^{d_1}\pi^{\frac{d_1}{2}}\Gamma\left(\frac{d_1}{2}\right)}
\sum_{l=0}^{\infty}\int_{m_l}^{\infty} x(x^2-m_l^2)^{\frac{d_1-2}{2}}\log\left\{1-\frac{(\beta_1 x-\alpha_1)(\beta_2x-\alpha_2)}{(\beta_1 x+\alpha_1)(\beta_2x+\alpha_2)}e^{-2ax}\right\}dx\\
&+\frac{aT^{\frac{d_1+1}{2}}}{2^{\frac{d_1-1}{2}}\pi^{\frac{d_1+1}{2}}}\sum_{l=0}^{\infty}\sum_{p=1}^{\infty}\left(\frac{m_l}{p}\right)^{\frac{d_1+1}{2}}K_{\frac{d_1+1}{2}}\left(\frac{p m_l}{T}\right)\\&-\frac{T^{\frac{d_1}{2}}}{2^{\frac{d_1-2}{2}}\pi^{\frac{d_1}{2}}}\sum_{l=0}^{\infty}\sum_{k=1}^{\infty}\sum_{p=1}^{\infty} \left(\frac{\sqrt{z_k^2+m_l^2}}{p}\right)^{\frac{d_1}{2}}
K_{\frac{d_1}{2}}\left(\frac{p}{T}\sqrt{z_k^2+m_l^2}\right).
\end{split}
\end{equation*}

The leading behavior of the finite temperature Casimir force acting on a pair of parallel plates when $R\ll a\ll r$ (i.e., the plate separation is much larger than the size of the extra dimensions, but much smaller than the size of the cross section) can be obtained from the corresponding  leading behavior of the Casimir force density acting on infinite parallel plates. We need to consider the case of high temperature and the case of low temperature   separately. In the high temperature regime, $aT\gg 1$.   The   leading term of the Casimir force when $R\ll a\ll r$ and $am\ll 1$ is obtained from the $l=p=0$ term in \eqref{eq7_27_4}:
\begin{equation*}
\begin{split}
F_{\text{Cas}}(a)\sim &-\frac{T\text{Vol}(\Omega)}{2^{d_1-2}\pi^{\frac{d_1-1}{2}}\Gamma\left(\frac{d_1-1}{2}\right)} \int_{m}^{\infty}\frac{\left(x^2- m^2\right)^{\frac{d_1-3}{2}}x^2dx}{\frac{(\beta_1 x+\alpha_1)(\beta_2x+\alpha_2)}{(\beta_1 x-\alpha_1)(\beta_2x-\alpha_2)}e^{2ax}-1}\\
\sim &-\frac{T\text{Vol}(\Omega)}{2^{d_1-2}\pi^{\frac{d_1-1}{2}}\Gamma\left(\frac{d_1-1}{2}\right)a^{d_1}} \int_{am}^{\infty}\frac{\left(x^2- (am)^2\right)^{\frac{d_1-3}{2}}x^2dx}{\frac{\left(\frac{\beta_1}{a} x+\alpha_1\right)\left(\frac{\beta_2}{a}x+\alpha_2\right)}{\left(\frac{\beta_1}{a} x-\alpha_1\right)\left(\frac{\beta_2}{a}x-\alpha_2\right)}e^{2x}-1}.
\end{split}
\end{equation*}If $\beta_1>0$ and $\beta_2>0$ (both non-Dirichlet conditions), then in the limit $a\ll \beta_1$ and $a\ll \beta_2$, we find that
the leading term is
\begin{equation}\label{eq7_29_11}
\begin{split}
F_{\text{Cas}}(a)\sim & -\frac{T\text{Vol}(\Omega)}{2^{d_1-2}\pi^{\frac{d_1-1}{2}}\Gamma\left(\frac{d_1-1}{2}\right)a^{d_1}} \int_{0}^{\infty}\frac{ x^{d_1-1} dx}{ e^{2x}-1}=-\frac{(d_1-1)\Gamma\left(\frac{d_1}{2}\right)\zeta_R(d_1)}{2^{d_1}\pi^{\frac{d_1}{2}}}\frac{T\text{Vol}(\Omega)}{a^{d_1}}.
\end{split}
\end{equation}This leading term is the same as for the case of $\beta_1=\beta_2=0$ (Dirichlet-Dirichlet boundary conditions). If $\beta_1=0$ and $\beta_2>0$ (one Dirichlet and one non-Dirichlet), then in the limit  $a\ll \beta_2$,
the leading term is
\begin{equation}\label{eq7_29_12}
\begin{split}
F_{\text{Cas}}(a)\sim & \frac{T\text{Vol}(\Omega)}{2^{d_1-2}\pi^{\frac{d_1-1}{2}}\Gamma\left(\frac{d_1-1}{2}\right)a^{d_1}} \int_{0}^{\infty}\frac{ x^{d_1-1} dx}{ e^{2x}+1}=\frac{(d_1-1)\Gamma\left(\frac{d_1}{2}\right)\zeta_R(d_1)}{2^{d_1}\pi^{\frac{d_1}{2}}}\left(1-2^{1-d_1}\right)\frac{T\text{Vol}(\Omega)}{a^{d_1}}.
\end{split}
\end{equation}Notice that if both plates have non-Dirichlet boundary conditions, the leading behavior is the same as both plates having Dirichlet boundary conditions. In this case, the Casimir force is attractive when $R\ll a\ll r$. On the other hand, if one plate assumes Dirichlet boundary condition and the other assumes non-Dirichlet boundary condition, then the leading behavior is the same as the Dirichlet-Neumann case, i.e., the Casimir force is repulsive in the limit $R\ll a\ll r$.

In the low temperature regime, i.e. $aT\ll 1$, the leading behavior   of the Casimir force when $R\ll a\ll r$ can be obtained analogously from \eqref{eq7_28_1}. We find  that when both plates assume non-Dirichlet boundary conditions or when both plates assume Dirichlet boundary conditions, the leading term is
\begin{equation}\label{eq7_29_13}
F_{\text{Cas}}(a)\sim  -\frac{d_1\Gamma\left(\frac{d_1+1}{2}\right)\zeta_R(d_1+1)}{2^{d_1+1}\pi^{\frac{d_1+1}{2}}}\frac{\text{Vol}(\Omega)}{a^{d_1+1}}.
\end{equation}When one plate assumes Dirichlet boundary condition and the other plate assumes non-Dirichlet boundary condition, the leading term is
\begin{equation}\label{eq7_29_14}
F_{\text{Cas}}(a)\sim  \frac{d_1\Gamma\left(\frac{d_1+1}{2}\right)\zeta_R(d_1+1)}{2^{d_1+1}\pi^{\frac{d_1+1}{2}}}(1-2^{-d_1})\frac{\text{Vol}(\Omega)}{a^{d_1+1}}.
\end{equation}
These zero temperature asymptotic behaviors have been observed in \cite{18}.

The asymptotics \eqref{eq7_29_11}, \eqref{eq7_29_12}, \eqref{eq7_29_13} and \eqref{eq7_29_14} are derived under the assumption that $R\ll a$, i.e. the size of the extra dimensions are much smaller than the plate separation. In the case $R\sim a$, one has to take into account the correction terms from the $l\neq 0$ terms in \eqref{eq7_27_4} and \eqref{eq7_28_1}. In the other extreme where $a\ll R$, the extra dimensions play the same role as the cross section in the macroscopic spacetime. Therefore, when $a\ll R$, the asymptotics of the Casimir force are obtained from \eqref{eq7_29_11}, \eqref{eq7_29_12}, \eqref{eq7_29_13} and \eqref{eq7_29_14} by replacing $d_1$ with $d=d_1+n$. We see that the sign of the Casimir force is not changed when we pass from $R\ll a$ to $a\ll R$, as long as $a\ll r$.

At first sight, it might be quite surprising to find that when the plate separation is small, the leading behavior of the Casimir force when non-Dirichlet boundary conditions are imposed on both plates is the same as when Dirichlet boundary conditions are imposed on both plates; but the leading behavior of the Casimir force when Dirichlet boundary condition is imposed on one plate and non-Dirichlet boundary condition is imposed on the other plate is the same as when Dirichlet boundary condition is imposed on one plate and Neumann boundary condition is imposed on the other. In fact, this can be explained as follows. The behavior of the Casimir force with respect to $a$ is governed by the solutions $z_k$ of the function $F(z)$ \eqref{eq3_27_3}, i.e., $z_k$ satisfies
\begin{equation*}
e^{2iaz_k}=\frac{(\alpha_1-i\beta_1 z_k)(\alpha_2-i\beta_2z_k)}{(\alpha_1+i\beta_1 z_k)(\alpha_2+i\beta_2z_k)}.
\end{equation*}When $k\rightarrow \infty$, $z_k\rightarrow \infty$. Therefore if $\beta_1>0$ and $\beta_2>0$,
\begin{equation*}
e^{2iaz_k} \sim 1\hspace{1cm}\text{as}\;\;k\rightarrow\infty.
\end{equation*}Consequently, we find that when $k$ is large enough,
\begin{equation*}
z_k \sim \frac{\pi k}{a}
\end{equation*}is close to the corresponding $z_k$ for the case where $\beta_1=\beta_2=0$. On the other hand, if $\beta_1=0$ and $\beta_2>0$, then when $k$ is large enough,
\begin{equation*}
e^{2iaz_k} \sim -1\hspace{1cm}\text{as}\;\;k\rightarrow\infty.
\end{equation*}This implies that
\begin{equation*}
z_k \sim \frac{\pi \left(k-1/2\right)}{a}
\end{equation*}is close to the corresponding $z_k$ for the case of Dirichlet-Neumann boundary conditions.

\section{Analysis of the sign of the Casimir force and its applications}\label{acf}In this section, we  analyze in more detail the sign of the Casimir force.
 The cases of Dirichlet-Dirichlet ($\beta_1=\beta_2=0$), Neumann-Neumann ($\alpha_1=\alpha_2=0$) or Dirichlet-Neumann ($\beta_1=0, \alpha_2=0$) boundary conditions have been studied in \cite{47}, where it was proved that for Dirichlet-Dirichlet or Neumann-Neumann case, the Casimir force is always attractive. For Dirichlet-Neumann case, the Casimir force is always repulsive.   In the following, we will not consider these cases.

First we consider the case $\beta_1=0$, $\alpha_2\neq 0$ and $\beta_2>0$ where Dirichlet boundary condition is imposed on one of the plates, and generic Robin boundary conditions on the other. In this case,
the finite temperature Casimir force between the plates is
\begin{equation}\label{eq7_28_4}\begin{split}
F_{\text{Cas}}(a; D;\alpha_2,\beta_2)=T\sum_{j=1}^{\infty}\sum_{l=0}^{\infty}\sum_{p=-\infty}^{\infty}\frac{\sqrt{m_{j,l}^2+[2\pi p T]^2}}{ \frac{ \beta_2 \sqrt{m_{j,l}^2+[2\pi p T]^2}+\alpha_2 }{   \beta_2\sqrt{m_{j,l}^2+[2\pi p T]^2} -\alpha_2}e^{2a\sqrt{m_{j,l}^2+[2\pi p T]^2}}+1},\end{split}
\end{equation}and the zero temperature Casimir force is
\begin{equation}\label{eq7_29_1}
\begin{split}
F_{\text{Cas}}^{T=0}(a;D;\alpha_2,\beta_2)
=&\frac{1}{\pi}\sum_{j=1}^{\infty}\sum_{l=0}^{\infty}\int_{m_{j,l}}^{\infty}\frac{x^2}{\frac{ \left( \beta_2 x+\alpha_2\right)}{ \left( \beta_2x-\alpha_2\right)}e^{2ax}+1}\frac{dx}{\sqrt{x^2-m_{j,l}^2}}.
\end{split}
\end{equation}\begin{figure}[h]
\epsfxsize=0.49\linewidth \epsffile{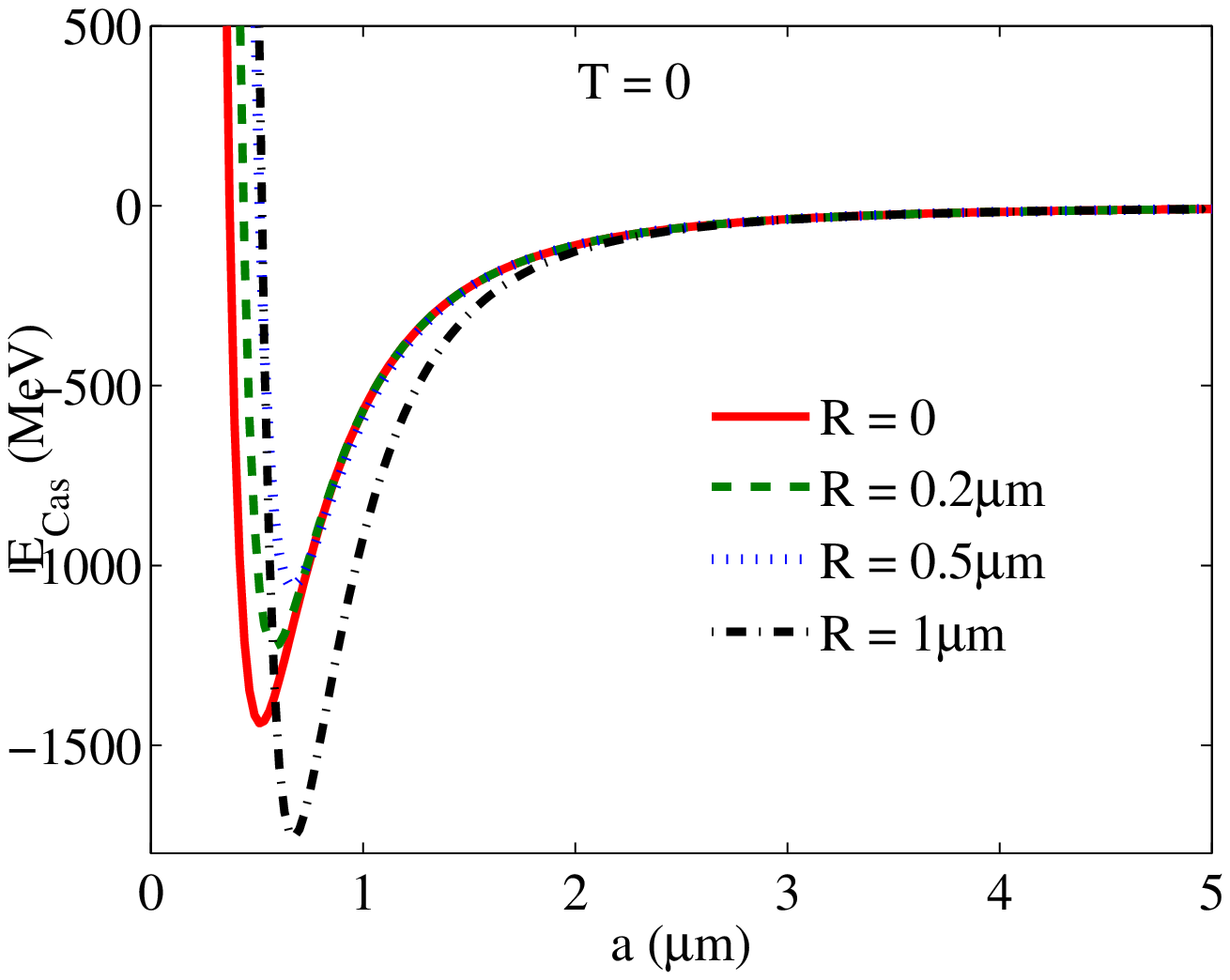} \epsfxsize=0.49\linewidth \epsffile{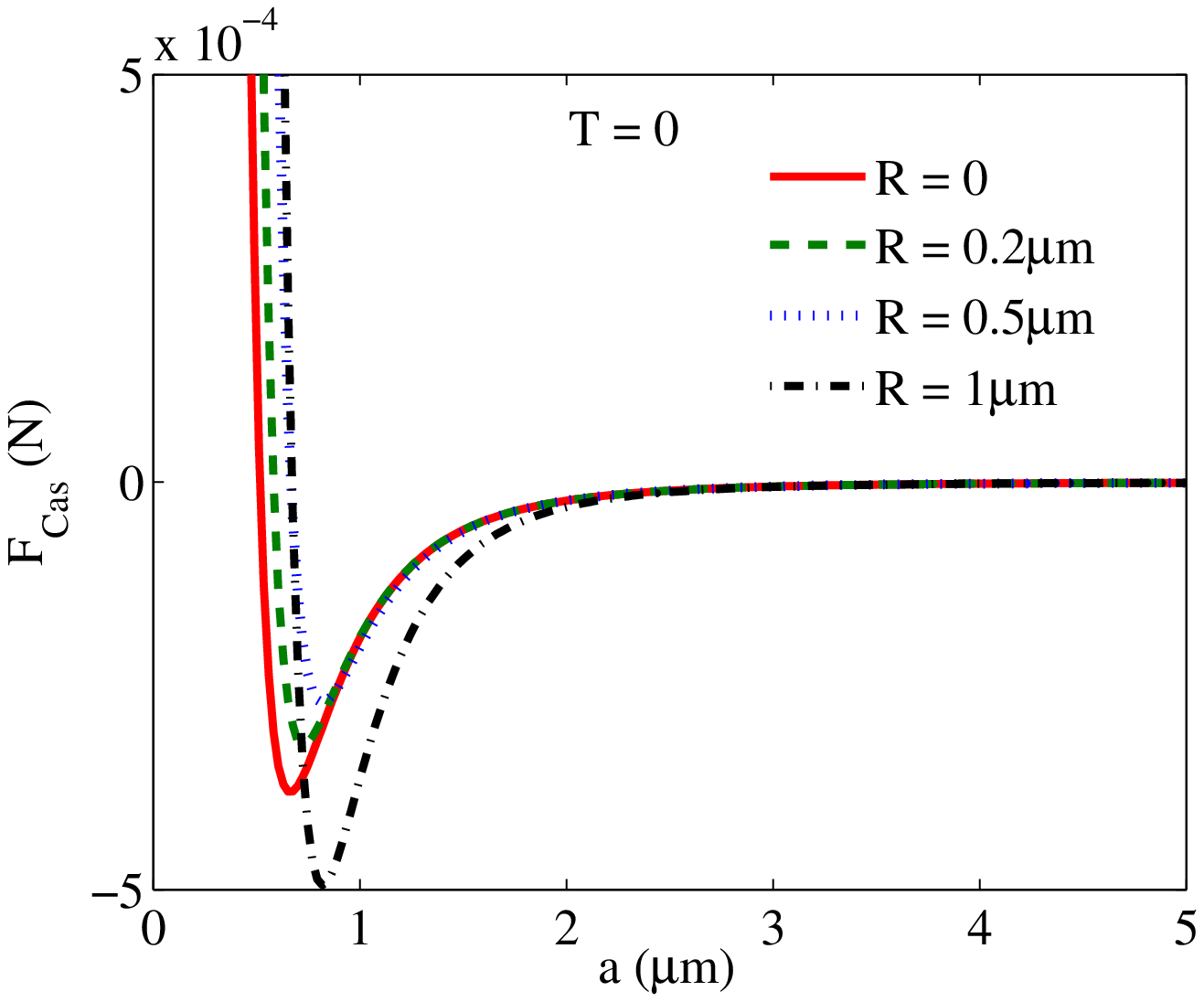}\caption{\label{f2} The Casimir energy $\Delta E_{\text{Cas}}(a)$ and Casimir force $F_{\text{Cas}}(a)$ due to a massless ($m=0$) scalar field when the macroscopic space is three dimension, i.e., $d_1=3$, in the presence of the internal manifold $S^1$ ($T^1$) with radius $R$. Here the cross section of the plates is a square $[0,L_2]\times [0,L_3]$ with $L_2=L_3=1\text{m}$. In this figure,  $\beta_1/\alpha_1=0\mu\text{m}$, $\beta_2/\alpha_2=0.3\mu\text{m}$ and $T=0$. }\end{figure}

\begin{figure}[h]
\epsfxsize=0.49\linewidth \epsffile{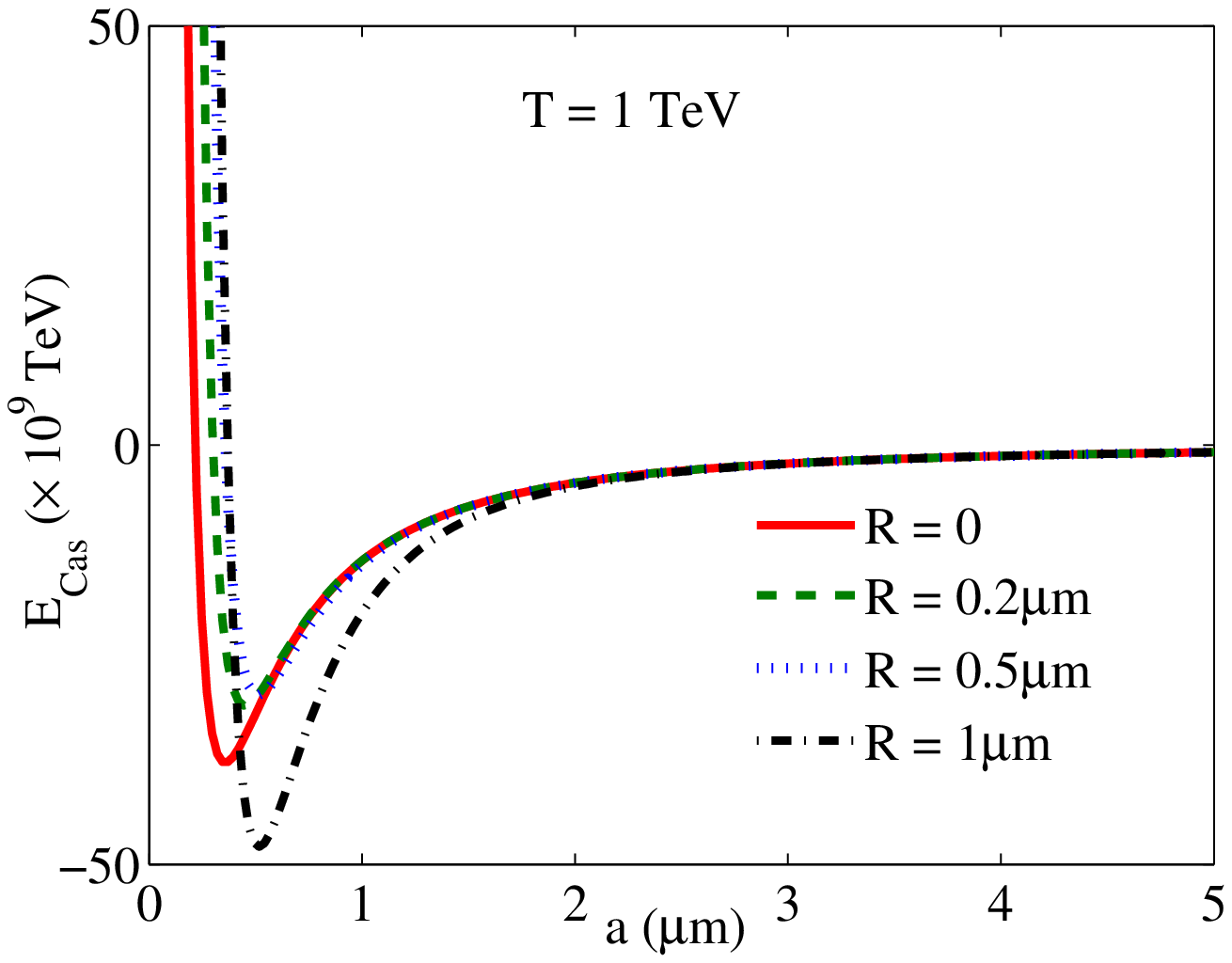} \epsfxsize=0.49\linewidth \epsffile{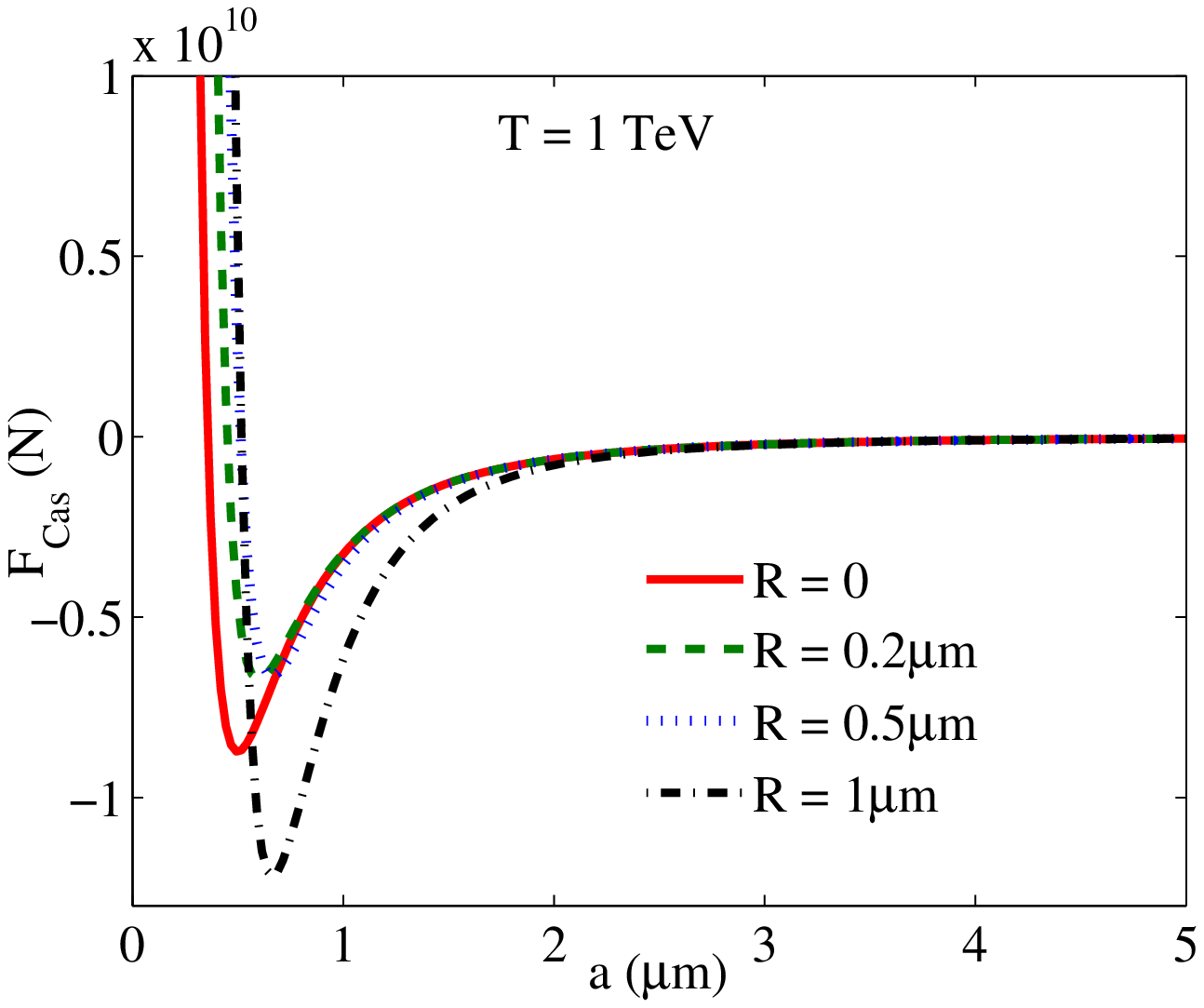}\caption{\label{f2_1} Same as FIG. \ref{f2} but with $T=1\,\text{TeV}$. }\end{figure}

If
\begin{equation}\label{eq7_23_9}
\begin{split}
 0<\frac{\alpha_2}{\beta_2}\leq   \min_{j,l,p}\left\{ \sqrt{m_{j,l}^2+[2\pi p T]^2}\right\}= \sqrt{\omega_{\Omega,1}^2+m^2},
\end{split}
\end{equation}then each term in the sum of \eqref{eq7_28_4} and \eqref{eq7_29_1} is positive. We find  that the Casimir force between the plates is always repulsive and the Casimir force is a monotonically decreasing function of $a$. Moreover, the presence of the extra dimensions enhances the Casimir force. The condition \eqref{eq7_23_9} means that the ratio $\alpha_2/\beta_2$ cannot be too large. In other words, this is a close-to-Neumann condition. Therefore, it is reasonable that the Casimir force is repulsive as in the Dirichlet-Neumann case. In general, we have shown in Section \ref{cf} that for any $\beta_2>0$,   the Casimir force is repulsive when $a\ll r$. When $aT\gg 1$, the leading term of the Casimir force is determined by the $p=l=0, j=1$ term in \eqref{eq7_28_4}, which is repulsive if \eqref{eq7_23_9} is satisfied, and attractive if \eqref{eq7_23_9} is not satisfied. At zero temperature, \eqref{eq7_29_1} also shows that if \eqref{eq7_23_9} is not satisfied, then when $am\gg 1$ or $a/r\gg 1$,   the Casimir force will eventually become attractive.   Therefore we see that if \eqref{eq7_23_9} is not satisfied, the Casimir force will change from repulsive to attractive at any temperature. In the massless case, the right hand side of \eqref{eq7_23_9} goes to zero when the size of the cross section $r$ goes to infinity. Therefore, for a pair of infinite parallel plates with Dirichlet boundary condition on one plate and non-Dirichlet and non-Neumann boundary condition on the other plate, the Casimir force always change from repulsive to attractive when $a$ increases from 0 to $\infty$.

\begin{figure}[h]
\epsfxsize=0.49\linewidth \epsffile{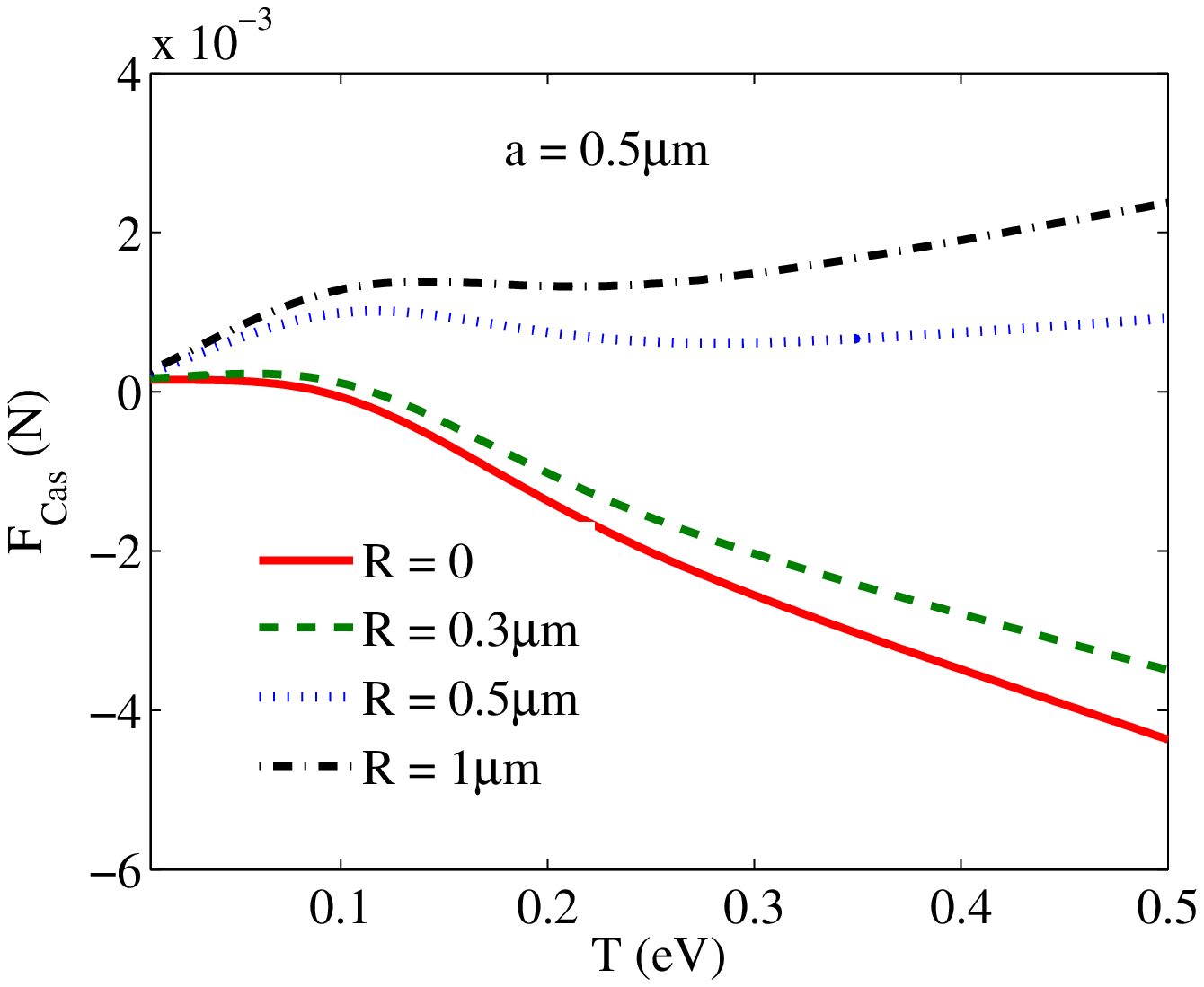} \epsfxsize=0.49\linewidth \epsffile{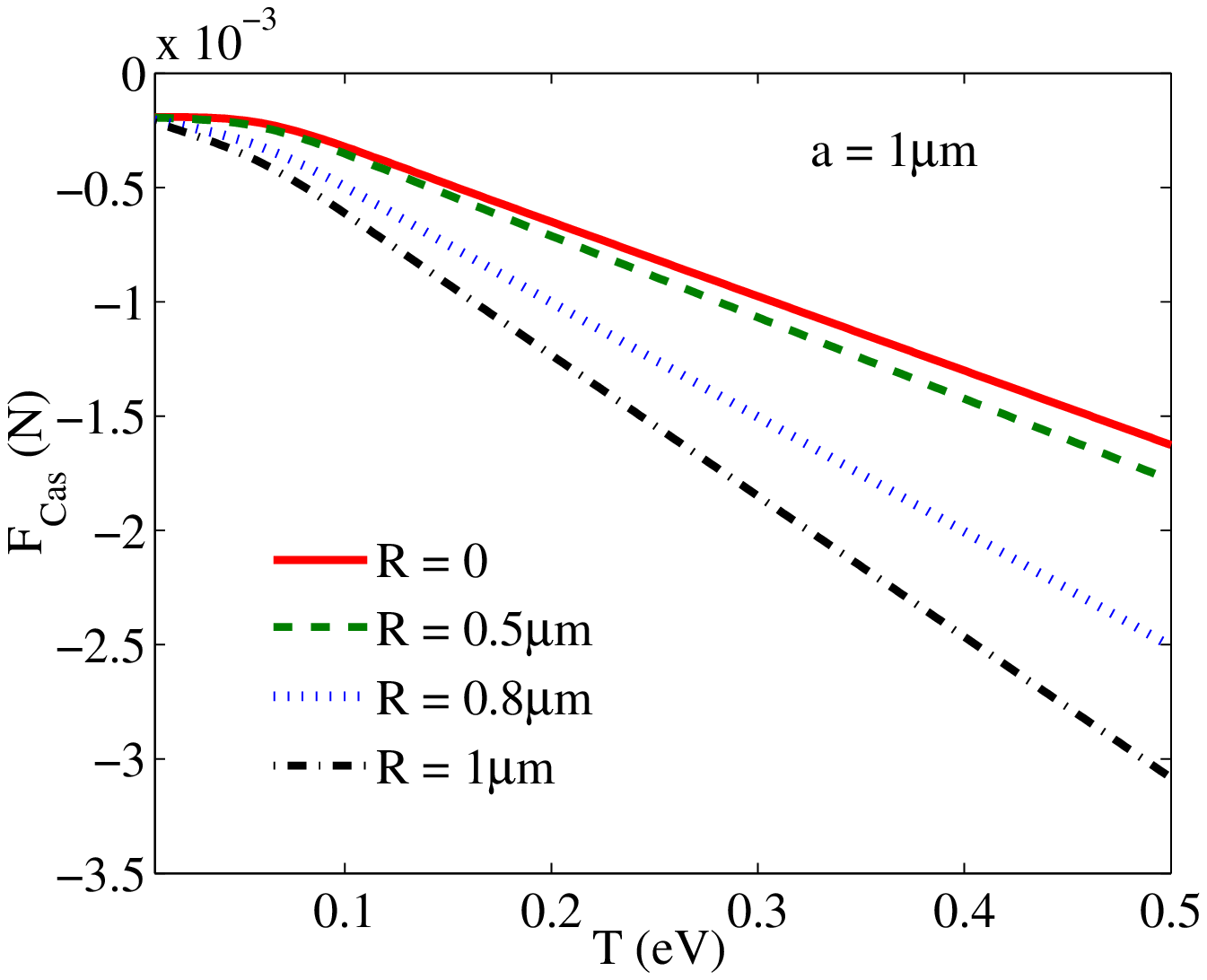}\caption{\label{f2_2} The dependence of the Casimir force on temperature when $a=0.5\mu$m and $a=1\mu$m. The other parameters are the same as in FIG. \ref{f2}. }\end{figure}

\begin{figure}[h]
\epsfxsize=0.49\linewidth \epsffile{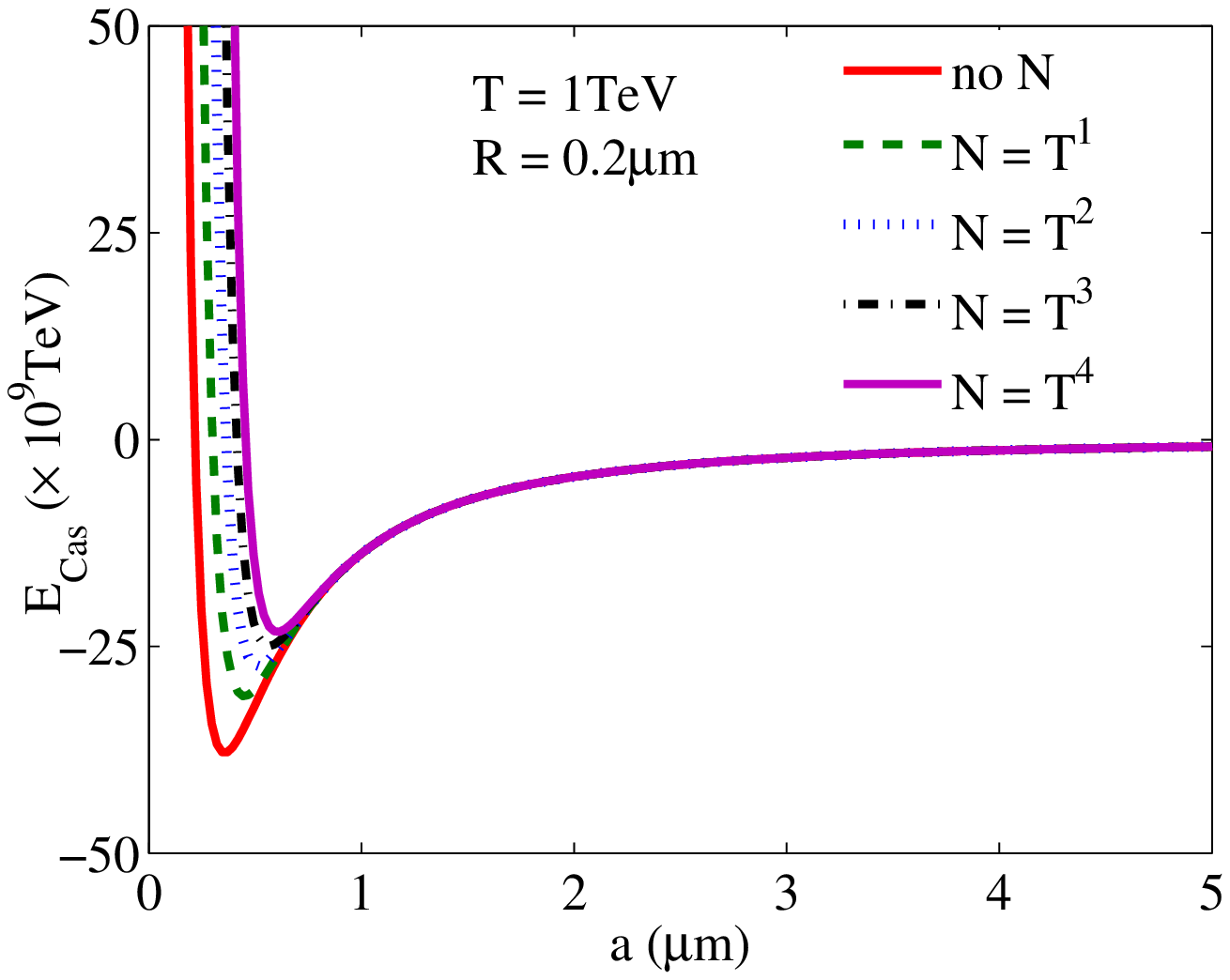} \epsfxsize=0.49\linewidth \epsffile{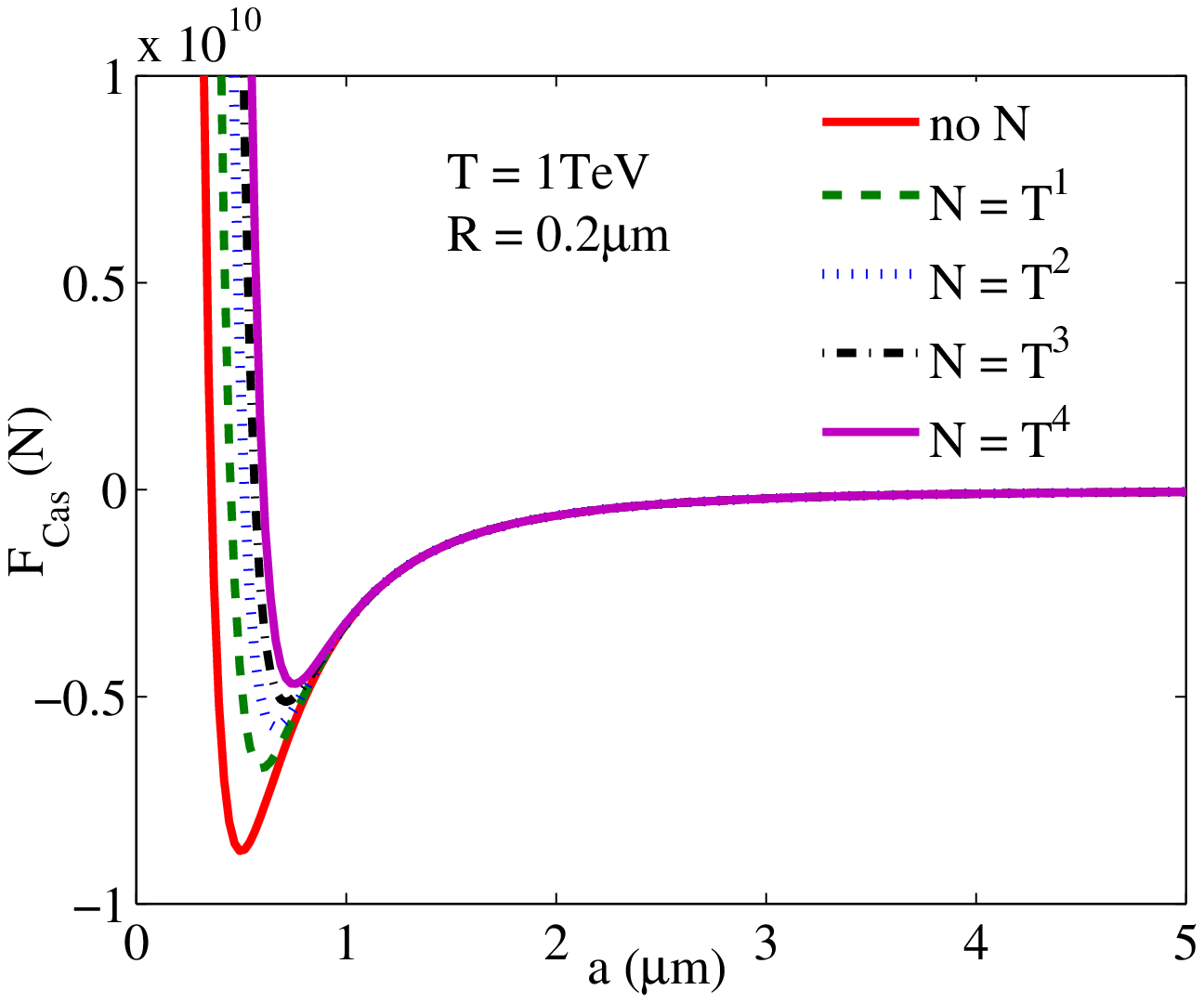}\caption{\label{f2_3} Same as FIG. \ref{f2_1}, but now the internal manifold is $T^1, T^2, T^3, T^4$, $T^n$ is a direct product of $n$ circles each with radius $R=0.2\mu$m. The other parameters are the same as in FIG. \ref{f2}. }\end{figure}

An example of the first case is shown in FIG. \ref{f2} and FIG. \ref{f2_1}, where the graphs for the Casimir energy and Casimir force acting on a pair of parallel plates embedded in an infinitely long rectangular cylinder with cross section $1\text{m}\times 1\text{m}$ due to massless scalar field is shown. The internal manifold is a circle with radius $R$. In these figures, we show the variation of the Casimir force with respect to the plate separation $a$ at $T=0$ and $T=1\,\text{TeV}$ ($1.16\times 10^{16}$\,K), in the case that the internal manifold has radius $R=0.2\mu$m, $R=0.5\mu$m and $R=1\mu$m and in the case without internal manifold $(R=0)$.  One of the plates assumes Dirichlet boundary condition, i.e., $\beta_1=0 $, and the other  assumes Robin boundary condition with  $\beta_2/\alpha_2=0.3\mu\text{m}$. Notice that $\alpha_2/\beta_2=3.33\times 10^{6}\text{m}^{-1} \geq \omega_{\Omega,1}=\pi \text{m}^{-1}$. These graphs show that the Casimir force is repulsive at small $a$ and becomes attractive for $a>a_c$. The critical point $a=a_c$ is a minimum point of the energy. Therefore, it is a stable equilibrium point. The position of the minimum point can be affected by the  size of the extra dimension and temperature.   The temperature dependence of the Casimir force when $a=0.5\mu$m and $a=1\mu$m are shown in FIG. \ref{f2_2}. They verify the linear dependence of the Casimir force when $aT>0.5$. Notice that with plate separation $a=1\mu$m,  the energy between the plates is of order    $10^{-3}$\,TeV at $T=0$ and of order $10^{10}$\,TeV at $T=1$\,TeV. The Casimir energy increases by a factor of $10^{13}$ from $T=0$ to $T=1$\,TeV. Therefore the temperature correction is important in the high temperature regime. Experiments on Casimir effect have not been able to reach this high energy regime. However, with the advent of technology, one can expect that future experiments will be able to explore the high temperature regime which might bring forward some new applications of Casimir effect in technology.    Return to FIG. \ref{f2_2}, we also notice that when the size of the extra dimension change from 0.2$\mu$m to $1\mu$m,   the Casimir force change from attractive to repulsive. As we have discussed in the previous section, this cannot happen for Dirichlet-Dirichlet, Neumann-Neumann or Dirichlet-Neumann boundary conditions. In FIG. \ref{f2_3}, we show the dependence of the Casimir force on plate separation when the internal manifold is $T^1=S^1$, $T^2, T^3$ and $T^4$ respectively, where $T^n$ is the product of $n$-circles each with radius $R=0.2\mu$m. This figure shows that contrary to the Dirichlet-Dirichlet, Neumann-Neumann or Dirichlet-Neumann cases, the increase in the number of extra dimensions can reduce the magnitude of the Casimir force.

The second case where $\alpha_1=0$, $\alpha_2\neq 0$ and $\beta_2>0$, i.e.,  Neumann boundary condition is imposed on one of the plates, and generic Robin boundary condition on the other is exactly the opposite of the first case. More precisely, in this case the finite temperature Casimir force and the zero temperature Casimir force are given respectively by
\begin{equation}\label{eq7_29_2}\begin{split}
F_{\text{Cas}}(a; N;\alpha_2,\beta_2)=-T\sum_{j=1}^{\infty}\sum_{l=0}^{\infty}\sum_{p=-\infty}^{\infty}\frac{\sqrt{m_{j,l}^2+[2\pi p T]^2}}{ \frac{ \beta_2 \sqrt{m_{j,l}^2+[2\pi p T]^2}+\alpha_2 }{   \beta_2\sqrt{m_{j,l}^2+[2\pi p T]^2} -\alpha_2}e^{2a\sqrt{m_{j,l}^2+[2\pi p T]^2}}-1},\end{split}
\end{equation}and
\begin{equation}\label{eq7_29_3}
\begin{split}
F_{\text{Cas}}^{T=0}(a;N;\alpha_2,\beta_2)
=&-\frac{1}{\pi}\sum_{j=1}^{\infty}\sum_{l=0}^{\infty}\int_{m_{j,l}}^{\infty}\frac{x^2}{\frac{ \left( \beta_2 x+\alpha_2\right)}{ \left( \beta_2x-\alpha_2\right)}e^{2ax}-1}\frac{dx}{\sqrt{x^2-m_{j,l}^2}}.
\end{split}
\end{equation}Therefore, we conclude immediately that when $\alpha_2/\beta_2$ is small enough to satisfy \eqref{eq7_23_9}, the Casimir force is always attractive as in the Neumann-Neumann case. However, if $\alpha_2/\beta_2$ does not satisfy \eqref{eq7_23_9}, the Casimir force is attractive at small plate separation, but will eventually   turn to repulsive when $aT$ or $a/r$ or $am$ is large enough. In the case \eqref{eq7_23_9} is satisfied, we can say more. As in the Dirichlet-Dirichlet or Neumann-Neumann case, \eqref{eq7_23_9} implies  that the magnitude of the Casimir force is a monotonically decreasing function of the plate separation. Moreover, since  each term in the sum of \eqref{eq7_29_2} and \eqref{eq7_29_3} is positive, and the sum of the $l=0$ terms in \eqref{eq7_29_2} and \eqref{eq7_29_3}  corresponds to the Casimir forces without extra dimensions, we find that when \eqref{eq7_23_9} is satisfied, the presence of extra dimensions enhances the magnitude of the Casimir force.

 \begin{figure}[h]
\epsfxsize=0.49\linewidth \epsffile{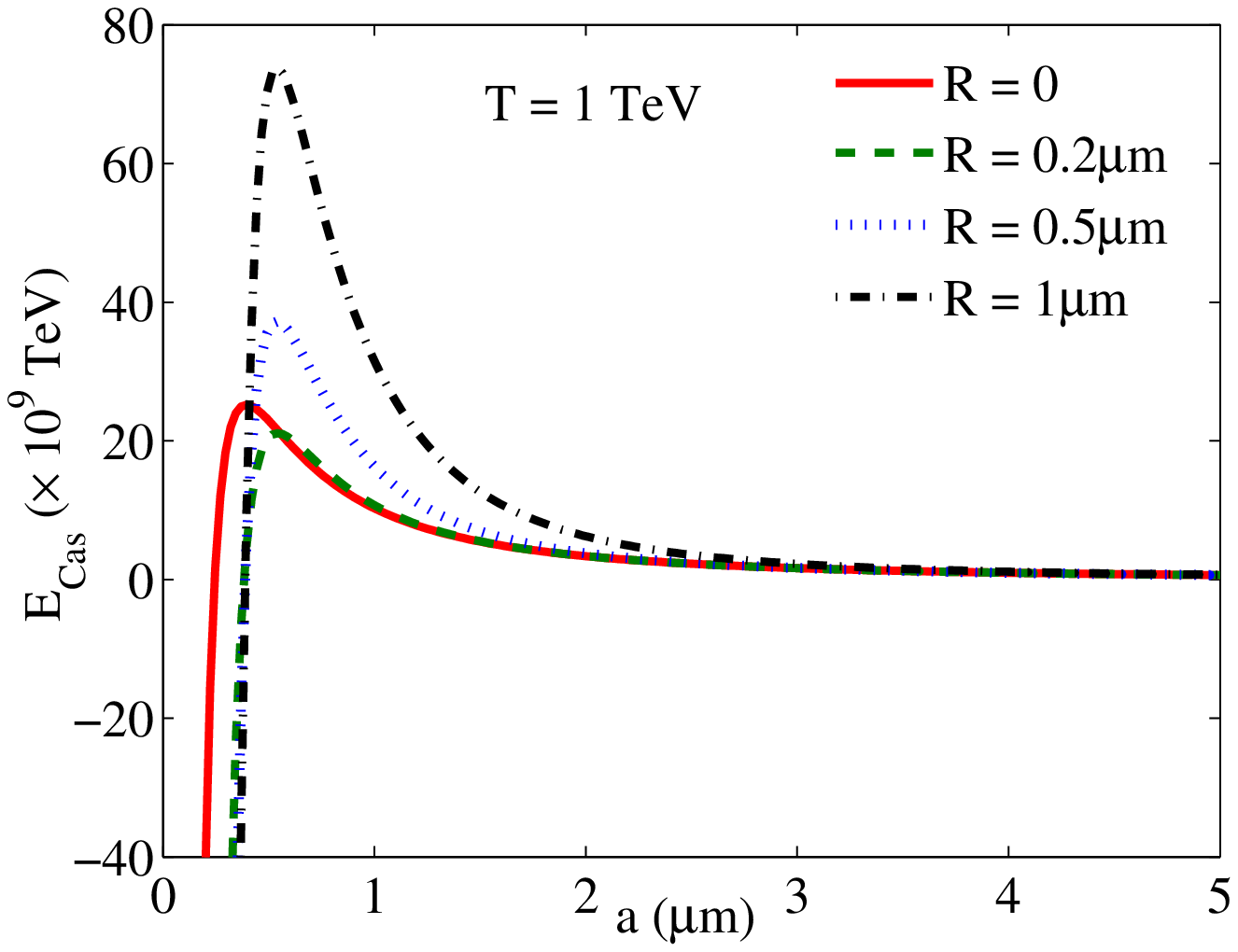}\epsfxsize=0.49\linewidth \epsffile{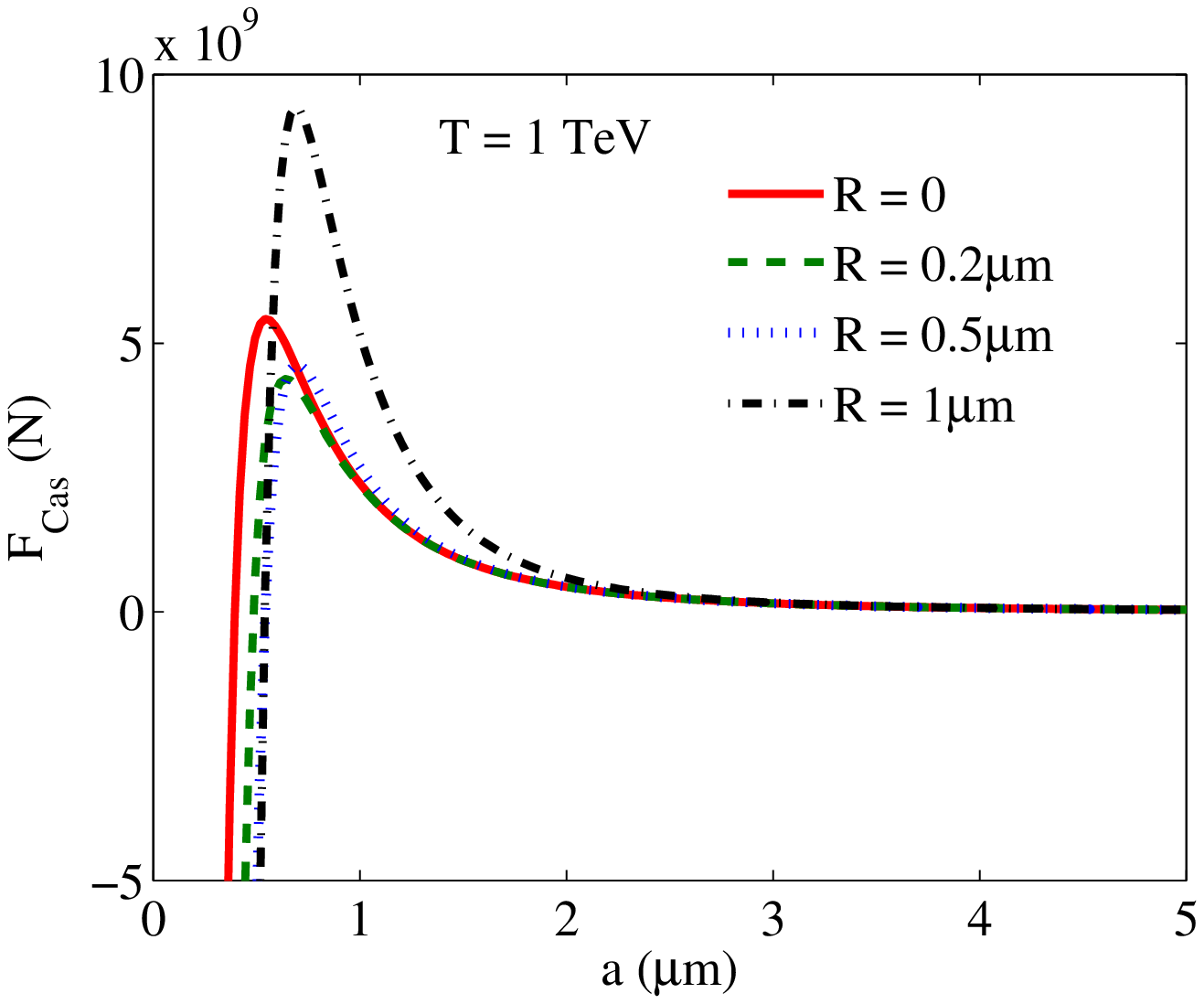} \caption{\label{f3}  Same as FIG. \ref{f2}, but with  $\beta_1/\alpha_1=0.4\text{m}$, $\beta_2/\alpha_2=0.3\mu\text{m}$ and $T=1\text{TeV}$. }\end{figure}

\begin{figure}[h]
\epsfxsize=0.49\linewidth \epsffile{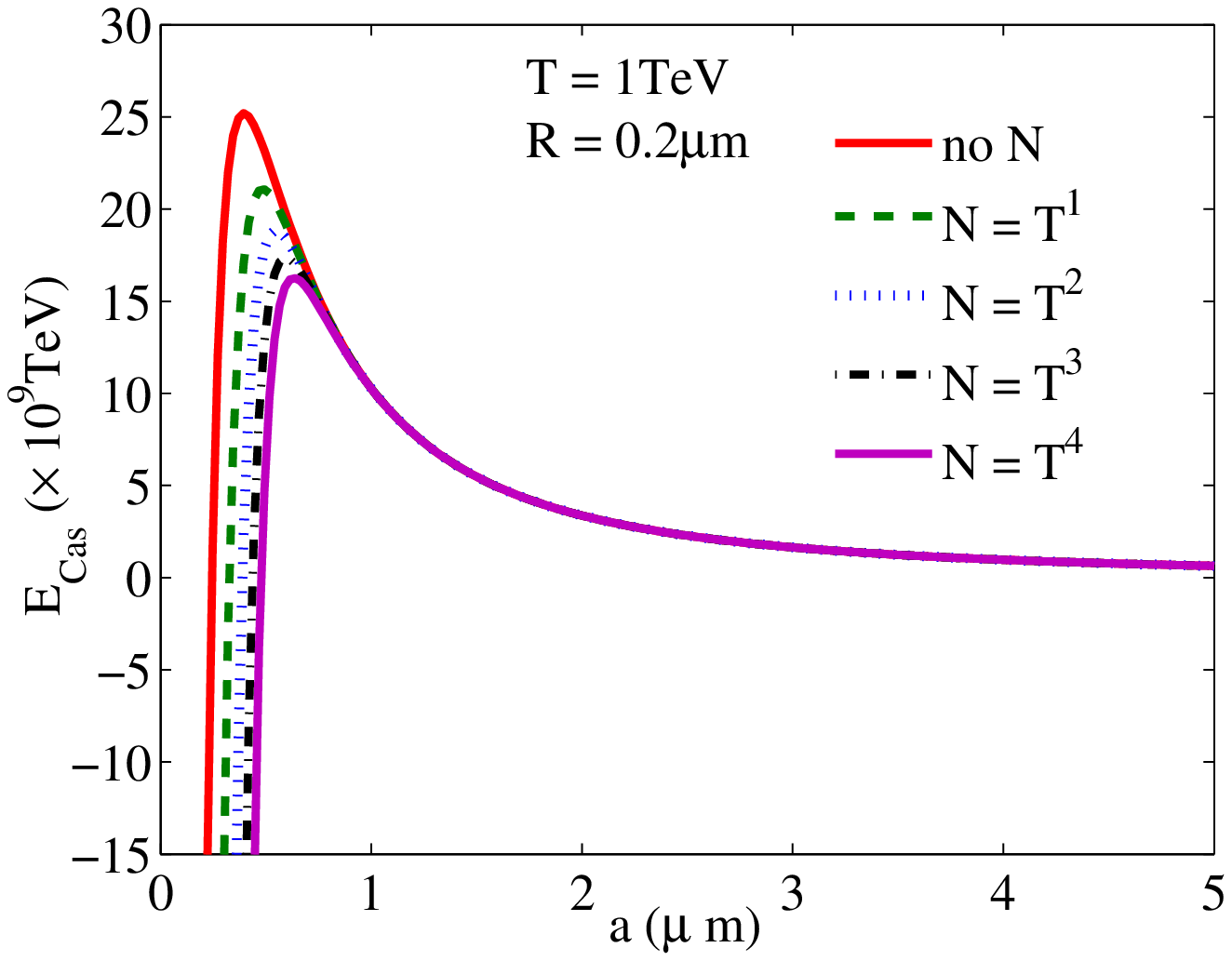} \epsfxsize=0.49\linewidth \epsffile{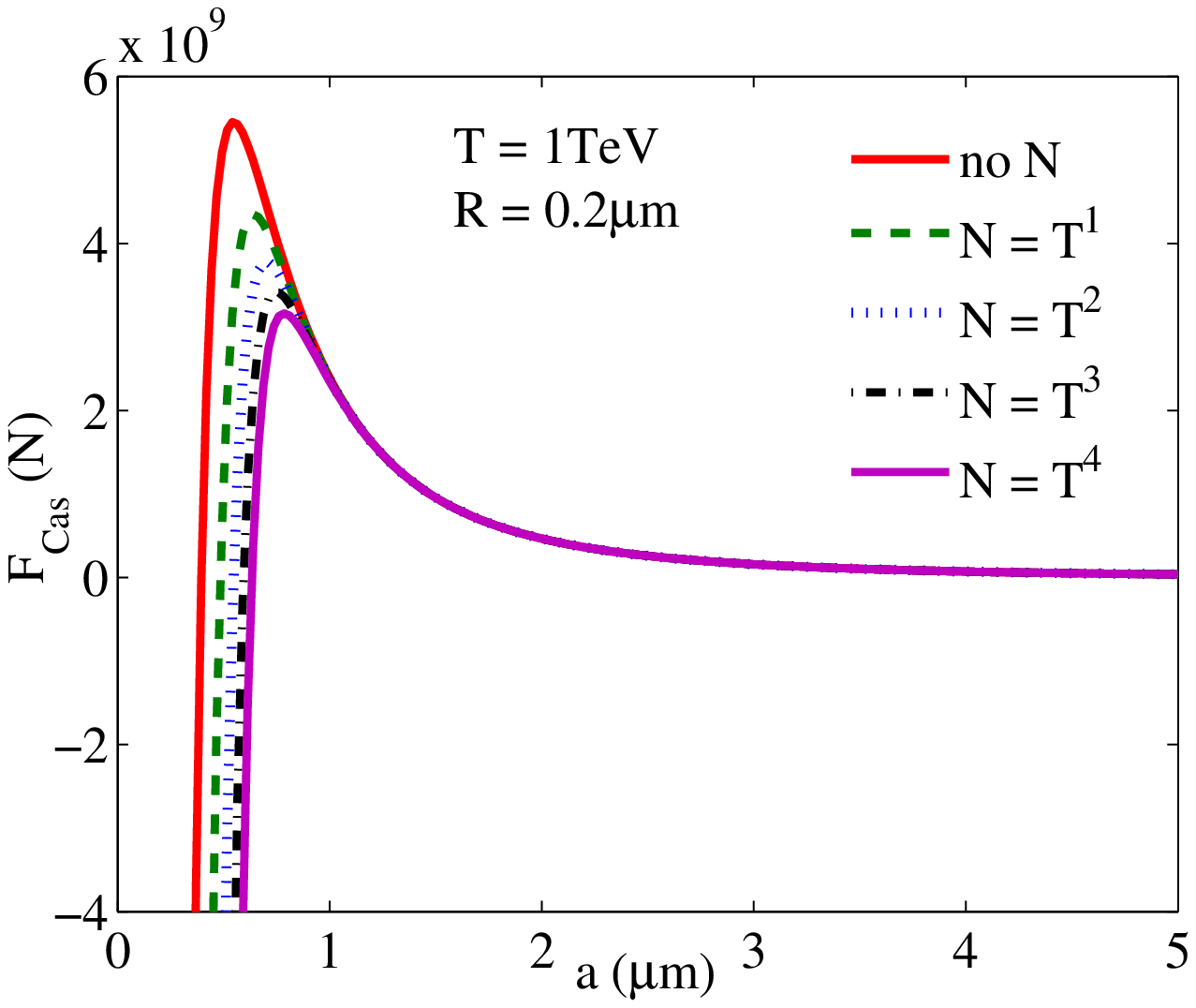}\caption{\label{f3_1} Same as FIG. \ref{f3}, but now the internal manifold is $T^1, T^2, T^3, T^4$, $T^n$ is a direct product of $n$ circles each with radius $R=0.2\mu$m. The other parameters are the same as in FIG. \ref{f3}. }\end{figure}

Finally, we consider the generic case where $\alpha_1>0,\beta_1>0$ and $\alpha_2>0, \beta_2>0$. Without loss of generality, assume that $\beta_2/\alpha_2\leq \beta_1/\alpha_1$.   As is already observed in Section \ref{cf}, when the Robin conditions on the two plates are the same, i.e., $\beta_1/\alpha_1=\beta_2/\alpha_2$, then the Casimir force is always attractive. If $\alpha_2/\beta_2$ satisfies \eqref{eq7_23_9}, so does $\alpha_1/\beta_1$. The Casimir force is then always attractive and its magnitude is enhanced by the presence of extra dimensions.  For any $\beta_1/\alpha_1>0$ and $\beta_2/\alpha_2>0$, we have   shown in Section \ref{cf} that the Casimir force is always attractive when the plate separation is small enough. In the other extreme where $a/r\gg 1$, if the temperature $T$ is not zero, the dominating term of the Casimir force is given by the term with $p=l=0$ and $j=1$ in \eqref{eq7_23_11}. We conclude that when $a$ is large enough, then the Casimir force is repulsive if \begin{equation*}
\frac{\alpha_1}{\beta_1} <\sqrt{\omega_{\Omega, 1}^2+m^2}<\frac{\alpha_2}{\beta_2},
\end{equation*}and is attractive otherwise. If the temperature $T$ is zero, the same conclusion can  be derived from \eqref{eq7_27_6}.

FIG. \ref{f3} and FIG. \ref{f3_1} show the dependence of the Casimir force on plate separation when $T =1\,\text{TeV}$. In FIG. \ref{f3}, the internal manifold is $S^1$ with different radius $R$. In FIG. \ref{f3_1}, the internal manifold is $T^1, T^2, T^3, T^4$ respectively.  The Robin coefficients in these graphs are $\beta_1/\alpha_1=0.4\text{m}$ and $\beta_2/\alpha_2=0.3\mu\text{m}$. Notice that $\alpha_1/\beta_1=2.5 \text{m}^{-1} < \pi \text{m}^{-1} < 3.33\times 10^6\text{m}^{-1}=\alpha_2/\beta_2\text{m}^{-1}$. The graphs show that the  Casimir  force is attractive at small $a$ and becomes repulsive at large $a$. There is only one equilibrium point which is unstable.

\begin{figure}[h]
\epsfxsize=0.49\linewidth \epsffile{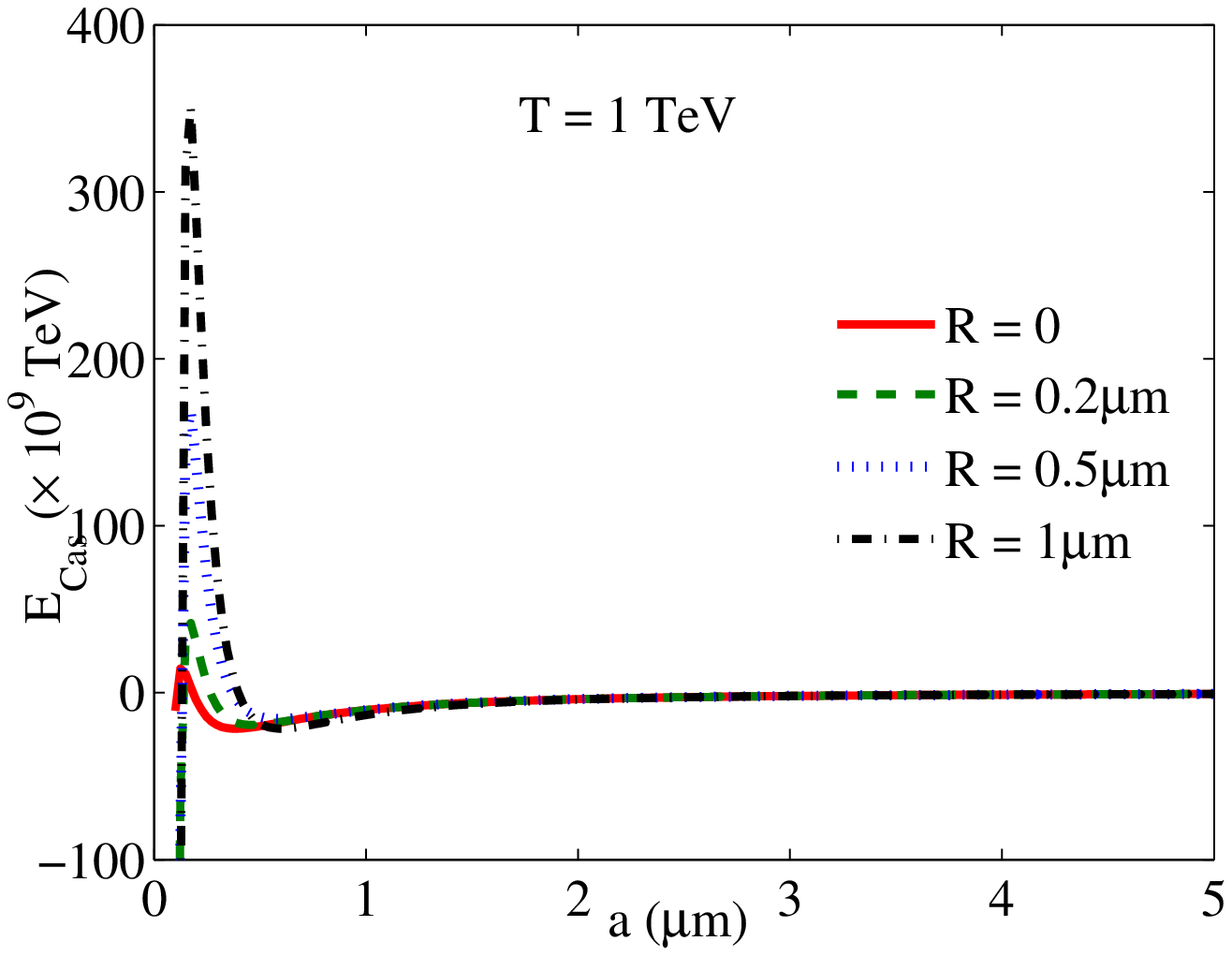}\epsfxsize=0.49\linewidth \epsffile{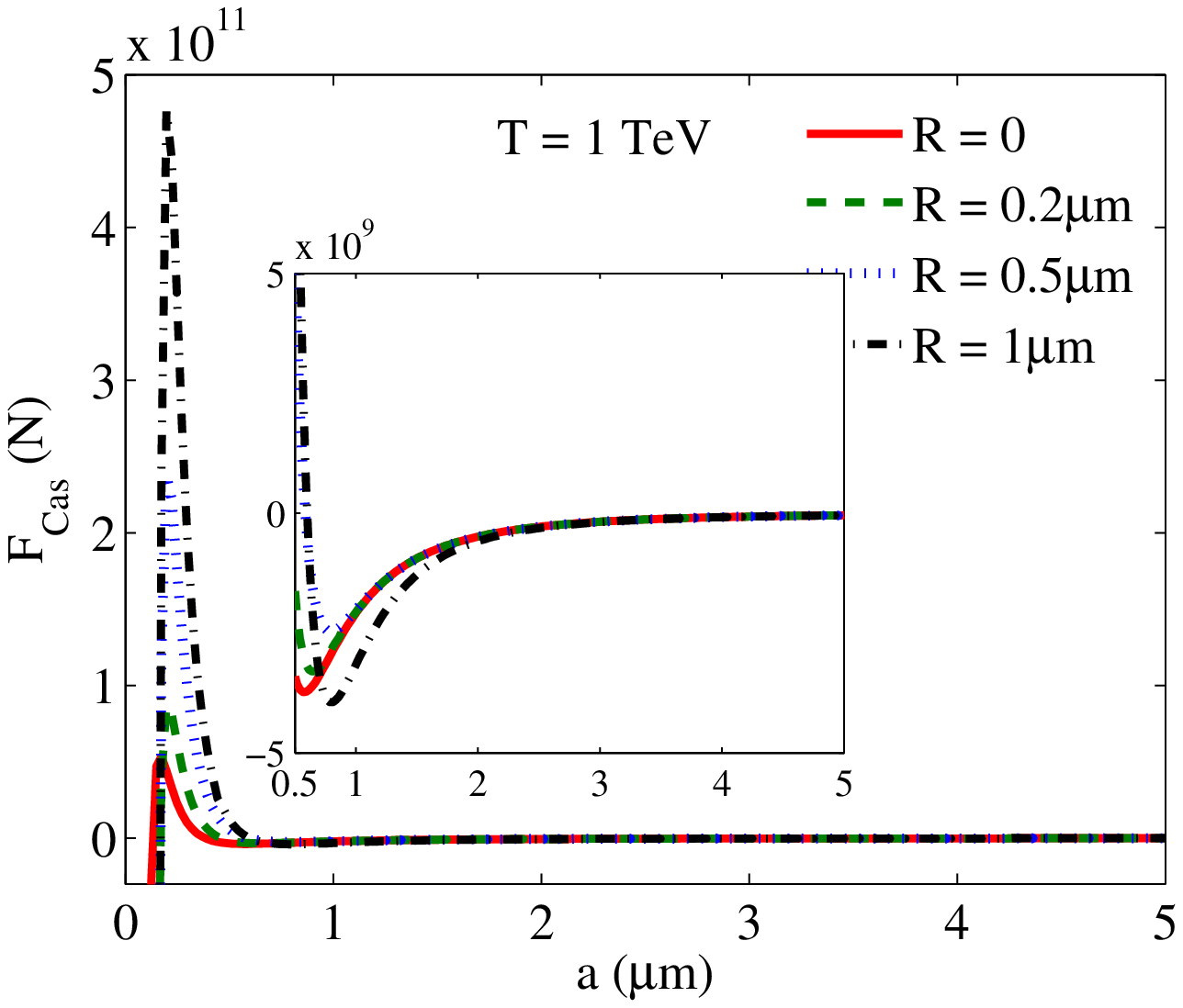} \caption{\label{f4}  Same as FIG. \ref{f2}, but with  $\beta_1/\alpha_1=0.08\mu\text{m}$, $\beta_2/\alpha_2=0.4\mu\text{m}$ and $T=1\text{eV}$. }\end{figure}

Another interesting case is shown in FIG. \ref{f4}, with $\beta_1/\alpha_1=0.08\mu\text{m}$ and $\beta_2/\alpha_2=0.3\mu\text{m}$. In this case, $\omega_{\Omega,1}<\alpha_1/\beta_1<\alpha_2/\beta_2$. The graph shows that the Casimir force is attractive at small and at large $a$ as dictated by our analysis above. However, the Casimir force can become repulsive at some intermediate values of $a$. This implies that there are two equilibrium points $a_1$ and $a_2$, one unstable and one stable. If the initial separation of the two plates is in the range of the two equilibrium points, i.e. $a_1<a<a_2$, then the two plates would tend to repulse each other until they settle at the distance $a=a_2$.

In the figures shown in this section, the size of the internal manifold is chosen to be between $0.2\mu$m to $1\mu$m so as to demonstrate significant difference with the Casimir force without extra dimensions. Extra dimensions of this size is not physically interesting. One can show that if the size of the internal manifold $R$ is ten times smaller than the plate separation, there is no significant difference between the Casimir force with or without the internal manifold. Therefore for physically interesting size of extra dimensions of order $10^{-12}$nm, it will not be detectable by the present experiments in Casimir effect which measures Casimir force  between objects  that are $10$nm $\sim 1000$nm apart.

Before ending this section, we would like to remark on the original setup we consider -- a piston moving freely inside a closed cylinder. As is shown by \eqref{eq7_29_4}, in this case, the Casimir force is the difference of the Casimir force between the left end of the cylinder and the piston, and the Casimir force between the right end of the cylinder and the piston. Since the magnitude of the Casimir force  is very large when the plate separation is very small, and is very small when the plate separation is very large, we can deduce that when the piston is close to one end, whether it is attracted to or pushed away from that end only depends on the Robin coefficients on the piston and on that end. However, when the piston is away from both ends, then which side it will move to depends on the Robin coefficients on the piston and on the two ends. There are some combinations of Robin coefficients that will make the piston stay in the middle region.

\section{Conclusion}
In this article, we studied the interplay between  geometry, temperature and boundary conditions on the sign and magnitude of the Casimir force acting on parallel plates. We first derived the finite temperature Casimir force acting on a  piston moving freely inside a closed  cylinder due to a scalar field with Robin boundary conditions. It is shown that even the Robin coefficients    on the two ends of the cylinder are different, the Casimir force acting on the piston is independent of the regularization procedure, since the divergent terms of the Casimir forces from  the two chambers of the cylinder divided by the piston are independent of the Robin coefficients and cancel each other.   By moving one end of the cylinder to infinity, we obtain the Casimir force acting on a pair of parallel plates.  In the high temperature regime, the  leading term of the Casimir force is linear in temperature, which shows that the   Casimir force has a classical limit. A modified Abel-Plana summation formula is used to rewrite the Casimir force which is suitable for the analysis of its low temperature behavior. When the parallel plates has finite size, the Casimir force decays exponentially when the temperature tends to zero. In case of infinite parallel plates, the temperature correction is of order $T^{d_1+1}$, where $d_1$ is the dimension of the macroscopic space. Interestingly, these behaviors are independent of the values of the Robin parameters.   The sign of the Casimir force when the plate separation is small and when the plate separation is large is analyzed in detail. It is found that if Dirichlet condition is imposed on one plate and non-Dirichlet condition is imposed on the other plate, then the Casimir force is repulsive when the plate separation is small enough. If non-Dirichlet conditions are imposed on both plates, the Casimir force is attractive for small enough plate separations. We give an explanation for these behaviors by the asymptotics of the frequencies. When the separation between the plates becomes sufficiently large, the sign of the Casimir force depends not only on the boundary conditions, but also on the geometry of the transversal dimensions. We show that for a wide range of Robin coefficients, the Casimir force can change from attractive to repulsive or repulsive to attractive, giving rise to unstable equilibrium and stable equilibrium respectively. This can be applied in nanotechnology if Robin conditions is used to model the skin depths of real materials.

As mentioned in the introduction, Robin boundary conditions arise naturally in Randall-Sundrum spacetime model. The results in this article is not readily transferred to the Randall-Sundrum model except for massless scalar field that couples   conformally to scalar curvature. There is a brief discussion of this in the zero temperature case in \cite{18}. It will be interesting to consider the finite temperature Casimir effect due to a bulk massive scalar field with general curvature coupling, and with general Robin boundary conditions on the branes. A special case has been considered in \cite{10_19_1} where the thermodynamic energy was shown to have a minimum that might give rise to brane stabilization mechanism.

\appendix
\section{The generalized Abel-Plana formula}\label{a1}
Here we present a more general Abel-Plana summation formula, which is a direct generalization of those presented in \cite{49,50}.
If $f_0(z), f_1(z)$ and $f_2(z) $ are meromorphic functions, and
\begin{equation}\label{eq6_24_2}\begin{split}
\lim_{Y\rightarrow \infty} \int_{b}^c \Bigl\{ f_0(x+iY)- f_1(x+ iY)\Bigr\}dx=0,\\
\lim_{Y\rightarrow \infty} \int_b^c \Bigl\{f_0(x-iY)- f_2(x- iY)\Bigr\}dx=0,
\end{split}
\end{equation} then
\begin{equation}\label{eq6_24_1}
\begin{split}
&\sum_{b\leq \text{Re}\; z\leq c} w_0(z) \text{Res}_{z}f_0(z) -\sum_{\substack{b\leq \text{Re}\; z\leq c\\ \text{Im}\;z\geq 0}} w_1(z) \text{Res}_{z}f_1(z)
 -\sum_{\substack{b\leq \text{Re}\; z\leq c\\ \text{Im}\;z\leq 0}} w_2(z) \text{Res}_{z}f_2(z) \\=&  \frac{1}{2\pi }\int_0^{\infty} \Bigl\{
\left. f_0(u+iy)-f_1(u+iy)\Bigr\}\right|_{u=b}^{u=c} dy+ \frac{1}{2\pi }\int_0^{\infty} \Bigl\{
\left. f_0(u-iy)-f_2(u-iy)\Bigr\}\right|_{u=b}^{u=c}dy-\frac{1}{2\pi i}\int_b^c \Bigl\{f_1(x)-f_2(x)\Bigr\}dx.
\end{split}
\end{equation}\begin{equation*}
\end{equation*}Here $w_0(z), w_1(z)$ and $w_2(z)$ are weight functions defined by\begin{equation*}\begin{split}w_0(z)&=\begin{cases}1, \hspace{0.5cm}&\text{if}\,\,z\in \mathfrak{D}_0, \\
 1/2, &\text{if}\,\,z\in\partial \mathfrak{D}_0,\end{cases},\hspace{1cm}\mathfrak{D}_0=\left\{ z\;:\;b<\text{Re}\,z<c\right\},\\ w_1(z) &=\begin{cases} 1, \hspace{0.5cm}&\text{if}\,\,z\in \mathfrak{D}_1,\\
   1/2, &\text{ if} \,\,z\in \partial \mathfrak{D}_1\setminus\{b,c\},\\
   1/4, &\text{if}\,\,z=b\,\,\text{or}\,\,c,\end{cases}\hspace{1cm}\mathfrak{D}_1=\left\{ z\;:\;b<\text{Re}\,z<c,\,\text{Im}z>0\right\}, \\w_2(z)&=\begin{cases}1, \hspace{0.5cm}&\text{if}\,\,z\in\mathfrak{D}_2,\\
     1/2, &\text{if}\,\,z\in\partial\mathfrak{D}_2\setminus\{b,c\},\\
   1/4, &\text{if}\,\,z=b\,\,\text{or}\,\,c,\end{cases}\hspace{1cm}\mathfrak{D}_2=\left\{ z\;:\;b<\text{Re}\,z<c,\,\text{Im}z<0\right\}.\end{split}\end{equation*}
This formula  can be proved in the same way as in \cite{49,50}. It is a direct consequence of the residue theorem.

To recover the original Abel-Plana summation formula, let $f(z)$ be a meromorphic function, and define
\begin{equation*}
\begin{split}
f_0(z)=& f(z)\frac{d}{dz}\log\left(e^{\pi i z}-e^{-i\pi z}\right)= i \pi f(z)\left(1+\frac{2}{e^{2\pi i z}-1}\right),\\f_1(z)=&f(z)\frac{d}{dz}\log \left(e^{-\pi i z}\right)=- i \pi f(z),\\
f_2(z)=&f(z)\frac{d}{dz}\log (e^{\pi i z})= i\pi f(z).
\end{split}
\end{equation*}If for all $x\geq 0$,
\begin{equation*}
\lim_{Y\rightarrow \infty}  f(x\pm iY)e^{-2\pi Y}=0,
\end{equation*}we can apply the  formula \eqref{eq6_24_1}, which gives
\begin{equation}\label{eq7_27_1}
\begin{split}
\frac{1}{2}f(0)+\sum_{p=1}^{\infty}f(p)=&\int_0^{\infty} f(x) dx +i\int_0^{\infty}\frac{f(iy)-f(-iy)}{e^{2\pi y}-1}dy +\pi i \sum_{ y>0} \frac{\text{Res}_{z=iy} f(z)-\text{Res}_{z=-iy}f(z)}{ e^{2\pi y}-1}\\&+2\pi i \sum_{\text{Re}\,z>0, \,\text{Im}\,z>0}\frac{\text{Res}_{z}f(z)}{e^{-2\pi i z}-1}-2\pi i \sum_{\text{Re}\,z>0, \,\text{Im}\,z<0}\frac{\text{Res}_{z}f(z)}{e^{2\pi i z}-1}.
\end{split}
\end{equation}Here we assume that $f(z)$ does not have poles at $z=n, n=0,1,2,\ldots$.  If $f(z)$ is analytic in the right-half plane, then the last three terms that contain the residues of $f$ in the right-half plane are identically zero. In this case, we obtain the original Abel-Plana summation formula.
\section{The zeta function $\zeta_{\text{cyl},T}(s)$}\label{a2}
 In this section, we want to compute the zeta function
\begin{equation*}
\begin{split}
\zeta_{\text{cyl},T}(s)=\sum_{k=1}^{\infty}\sum_{j=1}^{\infty}\sum_{l=0}^{\infty}\sum_{p=-\infty}^{\infty}\left(z_k^2+m_{j,l}^2+[2\pi pT]^2\right)^{-s},
\end{split}
\end{equation*}and its derivative at $s=0$. By definition, $z_k, k=1,2,\ldots$ are the  zeros of
\begin{equation*}
F(z)=(\alpha_1\alpha_2-\beta_1\beta_2z^2)\sin  az  +(\alpha_2\beta_1+\alpha_1\beta_2) z\cos az
\end{equation*} on the right half-plane. We can rewrite $F(z)$ as
\begin{equation*}
F(z)=\frac{1}{2i}\Bigl\{ (\alpha_1+i\beta_1z)(\alpha_2+i\beta_2z)e^{iaz}-(\alpha_1-i\beta_1z)(\alpha_2-i\beta_2z)e^{-iaz}\Bigr\}.
\end{equation*}Let $F_0(z)=F(z)$,
\begin{equation*}
F_1(z)=-\frac{1}{2i}(\alpha_1-i\beta_1z)(\alpha_2-i\beta_2z)e^{-iaz}, \hspace{0.5cm} F_2(z)=\frac{1}{2i}(\alpha_1+i\beta_1z)(\alpha_2+i\beta_2z)e^{iaz}.
\end{equation*}and define
\begin{equation*}
\begin{split}
f_i(z)= \sum_{j=1}^{\infty}\sum_{l=0}^{\infty}\sum_{p=-\infty}^{\infty}\left(z^2+m_{j,l}^2+[2\pi pT]^2\right)^{-s}\frac{d}{dz}\log F_i(z),\hspace{1cm}i=0,1,2.
\end{split}
\end{equation*}It is easy to verify that the conditions \eqref{eq6_24_2} are satisfied. Notice that $F_1(z)$ has zeros at $z=-i\alpha_j/\beta_j$, $j=1,2$, and $F_2(z)$ has zeros at $z=i\alpha_j/\beta_j$, $j=1,2$. For $i=0,1,2$, since $F_i(z)$ is a holomorphic function with simple zeros, the poles of $f_i(z)$ coincide with the zeros of $F_i(z)$.  Applying the generalized Abel-Plana summation formula \eqref{eq6_24_1}, we find that $\zeta_{\text{cyl},T}(s)$ can be written as a sum of three terms:
\begin{equation}\label{eq7_23_1}
\zeta_{\text{cyl},T}(s)=\zeta_{\text{cyl},T}^1(s)+\zeta_{\text{cyl},T}^2(s)+\zeta_{\text{cyl},T}^3(s).
\end{equation}The   term $\zeta_{\text{cyl},T}^1(s)$  is independent of $a$:
\begin{equation}\label{eq7_23_2}
\begin{split}
\zeta_{\text{cyl},T}^1(s)=&-\frac{1}{2}\zeta_{\Omega\times \mathcal{N},T}(s) +\sum_{i=1}^2 w_{i} \sum_{\substack{k,j\in\mathbb{N}, l\in\mathbb{N}_0, p\in\Z\\ m_{j,l}^2+[2\pi pT]^2\geq \left[\frac{\alpha_i}{\beta_i}\right]^2}}\left(-\left[\frac{\alpha_i}{\beta_i}\right]^2+m_{j,l}^2+[2\pi pT]^2\right)^{-s}\\&+\sum_{i=1}^2 w_{i} \sum_{\substack{k,j\in\mathbb{N}, l\in\mathbb{N}_0, p\in\Z\\ m_{j,l}^2+[2\pi pT]^2\leq \left[\frac{\alpha_i}{\beta_i}\right]^2}}\cos(\pi s)\left(\left[\frac{\alpha_i}{\beta_i}\right]^2-m_{j,l}^2-[2\pi pT]^2\right)^{-s}\\
&+\frac{1}{\pi}\int_0^{\infty} \sum_{j=1}^{\infty}\sum_{l=0}^{\infty}\sum_{p=-\infty}^{\infty}\left(x^2+m_{j,l}^2+[2\pi pT]^2\right)^{-s}\sum_{i=1}^2\frac{\alpha_i\beta_i}{\alpha_i^2+\beta_i^2x^2} dx.
\end{split}
\end{equation}The first term
\begin{equation*}
-\frac{1}{2}\zeta_{\Omega\times \mathcal{N},T}(s)=-\frac{1}{2}\sum_{j=1}^{\infty}\sum_{l=0}^{\infty}\sum_{p=-\infty}^{\infty}\left( m_{j,l}^2+[2\pi pT]^2\right)^{-s}
\end{equation*}
comes from the zero of $F_0(z)$ at $z=0$. If $\alpha_1=\alpha_2=0$, we have to change the sign of this term to positive.  The second and third terms in \eqref{eq7_23_2} come from the zeros of $F_1(z)$ and $F_2(z)$.  The weights $w_i$ are defined so that  if $\alpha_i=0$,  then  $w_i=1/2$; if $\alpha_i>0$, $\beta_i\geq 0$, then $w_i=0$ and if $\alpha_i> 0$, $\beta_i<0$,  then $w_i=1$. The last term in \eqref{eq7_23_2} comes from the $a$-independent part of
\begin{equation}\label{eq7_23_3}
-\frac{1}{2\pi i}\int_0^{\infty}(f_1(x)-f_2(x))dx.\end{equation}
The term $  \zeta_{\text{cyl},T}^2(s)$ is proportional to $a$, coming from the $a$-dependent part of \eqref{eq7_23_3}:
\begin{equation*}
\begin{split}
 \zeta_{\text{cyl},T}^2(s)=&\frac{a}{\pi}\int_0^{\infty} \sum_{j=1}^{\infty}\sum_{l=0}^{\infty}\sum_{p=-\infty}^{\infty}\left(x^2+m_{j,l}^2+[2\pi pT]^2\right)^{-s}dx\\
 =&\frac{a}{2\sqrt{\pi}}\frac{\Gamma\left(s-\frac{1}{2}\right)}{\Gamma(s)}\zeta_{\Omega\times\mathcal{N},T}\left(s-\frac{1}{2}\right).
\end{split}
\end{equation*}It is interesting to note that this term is independent of the Robin coefficients $\alpha_i,\beta_i$, $i=1,2$. Finally the first two terms on the right hand side of \eqref{eq6_24_1} give $\zeta_{\text{cyl},T}^3(s)$:
\begin{equation}\label{eq7_23_4}
\begin{split}
\zeta_{\text{cyl},T}^3(s) =\frac{1}{\pi}\sum_{j=1}^{\infty}\sum_{l=0}^{\infty}\sum_{p=-\infty}^{\infty}\int_{\sqrt{m_{j,l}^2+[2\pi p T]^2}}^{\infty}\sin (\pi s) \left(z^2-m_{j,l}^2-[2\pi p T]^2\right)^{-s} \frac{d}{dz}\log\left(1-\frac{(\beta_1 z-\alpha_1)( \beta_2z-\alpha_2)}{( \beta_1 z+\alpha_1)( \beta_2 z+\alpha_2)}e^{-2az}\right)dz.
\end{split}
\end{equation}This term goes to zero as $a\rightarrow \infty$.   From \eqref{eq7_23_4}, we find  that $\zeta_{\text{cyl},T}^3(0)=0$. Therefore $\zeta_{\text{cyl},T}(0)=\zeta_{\text{cyl},T}^1(0)+\zeta_{\text{cyl},T}^2(0)$ is linear in $a$. Moreover, the coefficient of $a$ is independent of the Robin coefficients.  The derivative of $\zeta_{\text{cyl},T}^3(s)$ at $s=0$ can be easily computed from \eqref{eq7_23_4} and we find that
\begin{equation}\label{eq7_24_5}\begin{split}
\zeta_{\text{cyl},T}'(0)=&\left(\zeta_{\text{cyl},T}^1\right)'(0)+\left(\zeta_{\text{cyl},T}^2\right)'(0)\\&-\sum_{j=1}^{\infty}\sum_{l=0}^{\infty}\sum_{p=-\infty}^{\infty}\log\left(
1-\frac{\left( \beta_1 \sqrt{m_{j,l}^2+[2\pi p T]^2}-\alpha_1\right)\left( \beta_2\sqrt{m_{j,l}^2+[2\pi p T]^2}-\alpha_2\right)}{\left( \beta_1 \sqrt{m_{j,l}^2+[2\pi p T]^2}+\alpha_1\right)\left( \beta_2 \sqrt{m_{j,l}^2+[2\pi p T]^2}+\alpha_2\right)}e^{-2a\sqrt{m_{j,l}^2+[2\pi p T]^2}}\right).
\end{split}\end{equation}
\section{The heat kernel coefficients $c_{\text{cyl},i}$ }\label{a3}In this section, we show that the heat kernel coefficients $c_{\text{cyl},i}$, $0\leq i\leq d+1$, \eqref{eq4_1_4} are linear functions of $a$. Moreover, the coefficients of $a$ is independent of the Robin coefficients $\alpha_i, \beta_i$, $i=1,2$.

 From the theory of elliptic operators, we have
\begin{equation*}
 \sum_{j=1}^{\infty}\sum_{l=0}^{\infty} e^{-t (
 \omega_{\Omega,j}^2+\omega_{\mathcal{N},l}^2)  }=\sum_{i=0}^{d+1} c_{\Omega\times\mathcal{N},i}t^{i-\frac{d-1}{2}}+O\left(t^{\frac{3}{2}}\right),\hspace{0.5cm}\text{as}\;\;t\rightarrow 0^+.
\end{equation*}
Using inverse Mellin transform, we find that\begin{equation}\label{eq7_23_7}\begin{split}
\sum_{k=1}^{\infty}  e^{-t( z_k^2+m^2) }=\frac{1}{2\pi i}\int_{\text{c}-i\infty}^{\text{c}+i\infty}  \Gamma(s) t^{-s} \zeta_{I}(s) ds,
\end{split}\end{equation}where
\begin{equation*}
\zeta_{I}(s)=\sum_{k=1}^{\infty} \left(z_k^2+m^2\right)^{-s}.
\end{equation*}As in Appendix \ref{a2}, we find that
\begin{equation*}
\zeta_{I}(s)=\zeta_{I}^1(s)+\zeta_{I}^2(s)+\zeta_{I}^3(s),
\end{equation*}where \begin{equation}\label{eq7_23_6}
\begin{split}
\zeta_{I}^1(s)=&-\frac{1}{2}m^{-2s} +\sum_{i=1}^2 w_{i}  \left(-\left[\frac{\alpha_i}{\beta_i}\right]^2+m^2\right)^{-s}\delta\left(m^2-\left[\frac{\alpha_i}{\beta_i}\right]^2\right) +\sum_{i=1}^2 w_{i} \cos(\pi s)\left(\left[\frac{\alpha_i}{\beta_i}\right]^2-m^2\right)^{-s}\delta\left(\left[\frac{\alpha_i}{\beta_i}\right]^2-m^2\right)\\
&+\frac{1}{\pi}\int_0^{\infty}  \left(x^2+m^2\right)^{-s}\sum_{i=1}^2\frac{\alpha_i\beta_i}{\alpha_i^2+\beta_i^2x^2} dx.
\end{split}
\end{equation}is independent of $a$,
\begin{equation*}
\zeta_{I}^2(s)=\frac{a}{2\sqrt{\pi}}\frac{\Gamma\left(s-\frac{1}{2}\right)}{\Gamma(s)}  m^{-2s+1}
\end{equation*}is proportional to $a$ and independent of the Robin coefficients, and
\begin{equation*}
\zeta_{I}^3(s) =\frac{1}{\pi}  \int_{ m}^{\infty}\sin (\pi s) \left(z^2-m^2 \right)^{-s} \frac{d}{dz}\log\left(1-\frac{(\beta_1 z-\alpha_1)( \beta_2z-\alpha_2)}{( \beta_1 z+\alpha_1)( \beta_2 z+\alpha_2)}e^{-2az}\right)dz
\end{equation*}is an analytic function of $s$ on the complex plane. Therefore, all the poles of the function $\zeta_{I}(s)$ come from $\zeta_{I}^1(s)+\zeta_{I}^2(s)$. It is easy to see that the poles of $\Gamma(s)\zeta_I^2(s)$ are at $s=1/2, -1/2, -3/2,-5/2, \ldots$ and all of them are simple poles. For $\zeta_I^1(s)$, the poles can only comes from the term
\begin{equation*}
\frac{1}{\pi}\int_0^{\infty}  \left(x^2+m^2\right)^{-s}\sum_{i=1}^2\frac{\alpha_i\beta_i}{\alpha_i^2+\beta_i^2x^2} dx=\frac{\beta_i}{\alpha_i}\frac{1}{\pi\Gamma(s)}\int_0^{\infty} t^{s-1}e^{-tm^2}\int_0^{\infty}\frac{e^{-tx^2}}{1+\left(\frac{\beta_i}{\alpha_i}\right)^2x^2}dxdt.\end{equation*}Now,
\begin{equation*}
\begin{split}
\int_0^{\infty}\frac{e^{-tx^2}}{1+\kappa x^2}dx=&\int_0^{\infty} e^{-u}\int_0^{\infty} e^{-(t+u\kappa)x^2}dxdu=\frac{\sqrt{\pi}}{2}\int_0^{\infty}\frac{e^{-u}}{\sqrt{t+u\kappa}}du=\frac{\sqrt{\pi}}{\kappa}\int_{\sqrt{t}}^{\infty}e^{-\frac{v^2-t}{\kappa}}dv\\
=&\frac{\sqrt{\pi}}{\kappa}e^{\frac{t}{\kappa}}\left(\frac{\sqrt{\pi\kappa}}{2}-\sum_{j=0}^{\infty}\frac{(-1)^j}{j!(2j+1)}\frac{t^{j+\frac{1}{2}}}{\kappa^j}\right).
\end{split}
\end{equation*}This shows that
\begin{equation*}
e^{-tm^2}\int_0^{\infty}\frac{e^{-tx^2}}{1+\kappa x^2}dx
\end{equation*}has an asymptotic expansion of the form
$
\sum_{j=0}^{\infty} c_j t^{\frac{j}{2}}
$ as $t\rightarrow 0^+$. It is then standard to show that the function $$\frac{\Gamma(s)}{\pi}\int_0^{\infty}  \left(x^2+m^2\right)^{-s}\sum_{i=1}^2\frac{\alpha_i\beta_i}{\alpha_i^2+\beta_i^2x^2} dx$$ has simple poles at $s=0, -1/2,-1, -3/2,-2, -5/2, \ldots$. It follows that the function $\Gamma(s)\zeta_I(s)$ only has simple poles at $s=1/2, 0, -1/2, -1, -3/2, \ldots$. Applying residue theorem to \eqref{eq7_23_7}, we find that
\begin{equation}\label{eq7_23_8}
\sum_{k=1}^{\infty}  e^{-t( z_k^2+m^2) }=\sum_{i=0}^{\infty} t^{i-\frac{1}{2}}\text{Res}_{s=\frac{1}{2}-i}\left(\Gamma(s)\zeta_I(s)\right)=\sum_{i=0}^{\infty} t^{i-\frac{1}{2}}\text{Res}_{s=\frac{1}{2}-i}\left(\Gamma(s)\zeta_I^1(s)\right)+\sum_{i=0}^{\infty} t^{i-\frac{1}{2}}\text{Res}_{s=\frac{1}{2}-i}\left(\Gamma(s)\zeta_I^2(s)\right).
\end{equation}Using the fact that
\begin{equation*}
\sum_{k=1}^{\infty}\sum_{j=1}^{\infty}\sum_{l=0}^{\infty} e^{-t\left(z_k^2+m_{j,l}^2 \right)}=\sum_{k=1}^{\infty}  e^{-t( z_k^2+m^2) }\sum_{j=1}^{\infty}\sum_{l=0}^{\infty} e^{-t (
 \omega_{\Omega,j}^2+\omega_{\mathcal{N},l}^2)  },
\end{equation*}we conclude from \eqref{eq7_23_8} that the heat kernel coefficients $c_{\text{cyl},i}$ are linear functions of $a$. Moreover, the coefficient of $a$ in $c_{\text{cyl},i}$ does not depend on the Robin coefficients $\alpha_i,\beta_i$, $i=1,2$ since $\zeta_I^2(s)$ does not.

\begin{acknowledgments}
This project is   funded by Ministry of Science, Technology and Innovation, Malaysia under e-Science fund 06-02-01-SF0080. We are grateful to A. Flachi and the anonymous referee for the helpful comments and suggestions.
\end{acknowledgments}

\end{document}